\documentclass[12pt]{article}
\usepackage{epsfig}
\usepackage{rotating}

\catcode`\@=11
\DeclareMathAlphabet{\mathpzc}{OT1}{pzc}{m}{it}

\newcommand{\insertfig}[2]{\mbox{\epsfxsize=#1cm \epsfbox{#2.eps}}}

\newcommand{\cQ}{{\cal Q}}
\newcommand{\Bx}{x_{\rm B}}

\font\cmss=cmss12 
\def\1{\hbox{{1}\kern-.25em\hbox{l}}}
\def\bfZ{\relax{\hbox{\cmss Z\kern-.4em Z}}}

\def \be  {\begin{equation}}
\def \ee  {\end{equation}}
\def \ba  {\begin{eqnarray}}
\def \ea  {\end{eqnarray}}
\def \baa {\begin{eqnarray*}}
\def \eaa {\end{eqnarray*}}
\def \bb  {\begin {thebibliography} }
\def \eb  {\end{thebibliography}}
\def \lab #1 {\label{#1}}

\newcommand\re[1]{(\ref{#1})}

\def \matrix #1 {\left(\begin{array}{cc} #1 \end{array}\right)}

\newcommand{\as}{\ifmmode\alpha_{\rm s}\else{$\alpha_{\rm s}$}\fi}
\newcommand{\asbar}{\ifmmode\bar{\alpha}_{\rm s}\else{$\bar{\alpha}_{\rm s}$}\fi}

\newcommand{\bit}[1]{\mbox{\boldmath$#1$}}
\newcommand{\ft}[2]{{\textstyle\frac{#1}{#2}}}

\font\cmss=cmss12 
\def\inbar{\,\vrule height1.5ex width.4pt depth0pt}
\def\IC{\relax\hbox{$\inbar\kern-.3em{\rm C}$}}
\def\IZ{\relax{\hbox{\cmss Z\kern-.4em Z}}}
\def\IR{{\hbox{{\rm I}\kern-.2em\hbox{\rm R}}}}

\def\IP{{\hbox{{\rm I}\kern-.2em\hbox{\rm P}}}}
\def\II{\hbox{{1}\kern-.25em\hbox{l}}}

\def\numberbysection{\@addtoreset{equation}{section}
                     \def\theequation{\thesection.\arabic{equation}}}
\numberbysection

\newbox\lett\newdimen\lheight\newdimen\lwidth
\def\ontop#1#2{\setbox\lett=\hbox{#2}\lheight\ht\lett
\multiply\lheight by 12 \divide\lheight by 10\relax%
\lwidth\wd\lett \multiply\lwidth by 8 \divide\lwidth by 10\relax #2\kern-\lwidth%
\raise\lheight\hbox{{$\scriptstyle #1$}}\kern.1ex}

\def\XXint#1#2#3{{\setbox0=\hbox{$#1{#2#3}{\int}$}
     \vcenter{\hbox{$#2#3$}}\kern-.5\wd0}}

\begin{document}

\begin{titlepage}

\thispagestyle{empty}

\vskip2cm

\centerline{\large \bf Exclusive electroproduction revisited:}

\vspace{2mm}

\centerline{\large \bf treating kinematical effects}

\vspace{15mm}

\centerline{\sc A.V. Belitsky$^a$, D. M\"uller$^b$}

\vspace{15mm}

\centerline{\it $^a$Department of Physics, Arizona State University}
\centerline{\it Tempe, AZ 85287-1504, USA}

\vspace{5mm}

\centerline{\it $^b$Institut f\"ur Theoretische Physik II, Ruhr-Universit\"at Bochum}
\centerline{\it D-44780 Bochum, Germany}

\vspace{3cm}

\centerline{\bf Abstract}

\vspace{5mm}

Generalized parton distributions of the nucleon are accessed via exclusive leptoproduction
of the real photon. While earlier analytical considerations of phenomenological observables
were restricted to twist-three accuracy, i.e., taking into account only terms suppressed by
a single power of the hard scale, in the present study we revisit this differential cross
section within the helicity formalism and restore power-suppressed effects stemming from the
process kinematics exactly. We restrict ourselves to the phenomenologically important case of
lepton scattering off a longitudinally polarized nucleon, where the photon flips its helicity
at most by one unit.

\end{titlepage}

\setcounter{footnote} 0

\newpage

\pagestyle{plain}
\setcounter{page} 1

\section{Electroproduction observables}
\label{Sec-AziAngDep}

Unravelling nucleon's structure from generalized parton distributions (GPDs)
\cite{MueRobGeyDitHor94} requires their measurements in exclusive leptoproduction
experiments. Recent years had witnessed groundbreaking efforts which put the
underlying theoretical framework on a firm basis with accuracy of approximation
involved being under control (see, e.g., \cite{Rev04}). While to date reliable
modeling of partonic correlations encoded in GPDs is far from being mature enough,
theoretical analyses of experimental observables are not constrained by any
complications of principle and rather awaited the time when experiments reached
competing precision.

The cross section for exclusive electroproduction of photons, being the cleanest probe
of GPDs, was computed analytically already for some time to twist-three accuracy
\cite{BelMueKir01}, i.e., keeping terms suppressed at most by one power of the hard
scale and neglecting everything else. While this approximation is robust for kinematical
regimes with moderately hard virtualities of the exchanged photon at large energies, it
was shown to overestimate available data at low momentum transfer in the valence quark region,
i.e., for moderate values of the Bjorken variable. This calls for the restoration of
contributions ignored previously on the basis of their parametric suppression. In a more
recent investigation \cite{BelMul09}, we demonstrated that the deviation between the data
and theoretical estimates could be reconciled by calculating kinematical corrections in
hard scale exactly while ignoring dynamical high-twist contributions altogether. The latter
assumptions can be motivated by the expected hierarchy of low-energy scales associated with
hadronic matrix elements of high-twist operators which are smaller than other soft kinematical
scales in the problem, like the hadron mass or the $t$-channel momentum transfer. This
phenomenon exhibits itself as precocious scaling in conventional deep-inelastic scattering.

While our earlier analysis was preformed for the (pseudo) scalar target \cite{BelMul09},
presently we will generalize this consideration to the case of a spin one-half hadron. The
main focus of our consideration is the differential cross section for scattering of the
electron/positron $\ell = e^\mp$ off the nucleon $N$ with the emission of the real photon in
the final state, $\ell (k) N (p_1) \to \ell (k^\prime) N (p_2) \gamma (q_2)$,
\be
\label{WQ}
d \sigma
=
\frac{\alpha^3  \Bx y^2 } {8 \, \pi \,  {\cal Q}^4 \sqrt{1 + \epsilon^2}}
\left| \frac{\cal T}{e^3} \right|^2
d \Bx d {\cal Q}^2 d |t| d \phi
\, .
\ee
The phase space of the process is parameterized by the Bjorken variable $\Bx = \mathcal{Q}^2/(2 p_1 \cdot q_1)$
determined in terms of the momentum $q_1 = k - k'$ carried by the virtual photon of mass $\mathcal{Q}^2 = -
q_1^2$, the square of the $t$-channel momentum $t \equiv \Delta^2$ with $\Delta = p_2 - p_1$ and the lepton
energy loss $y = p_1\cdot q_1/p_1\cdot k$. The azimuthal angle $\phi$ of the recoiled nucleon is defined in
the rest frame of the target with the $z$-axis directed counter-along the photon three-momentum $\bit{q}_1$.
While the theoretical analysis of the microscopic physics is cleanest when one formally takes the limit
$\mathcal{Q} \to \infty$, realistic experiments are done in a few GeV region where the effects from
kinematical parameters suppressed by $\mathcal{Q}$,
\be
\epsilon \equiv 2 \Bx \frac{M}{{\cal Q}}
\, , \qquad
\frac{t}{\mathcal{Q}^2}
\, ,
\ee
may be significant.

The electroproduction amplitude $\mathcal{T}$ is a linear superposition of the Bethe-Heitler
and deeply virtual Compton scattering (DVCS) amplitudes. In the former process, the real photon
is emitted from the lepton which then scatters off the target nucleons via the transition matrix
element of the electromagnetic quark current $J_\mu$, parameterized in terms of the Pauli and
Dirac form factors $F_1 = F_1 (t)$  and $F_2 = F_2 (t)$,
\be
\label{EMcurrentFFs}
J_\mu = \langle p_2 | j_\mu (0) | p_1 \rangle
=
\bar{u}_2 \left( \gamma_\mu F_1 +  i \sigma_{\mu\nu} \frac{\Delta^\nu}{2 M} F_2 \right) u_1
\, ,
\ee
with the nucleon bispinors $u_i = u (p_i)$ normalized conventionally as $\bar u u = 2 M$. The DVCS
amplitude
\be
\label{Def-Tensor-Had}
T_{\mu\nu} =\frac{ i}{e^2} \int d^4 z \, {\rm e}^{\frac{i}{2} (q_1 + q_2) \cdot z}
\langle p_2 | T \{ j_\mu (z/2) j_\nu (- z/2) \} | p_1 \rangle
\, ,
\ee
encodes the partonic structure of the nucleon and is the object of interest. In the square of
the scattering amplitude
\begin{equation}
{\cal T}^2
= |{\cal T}^{\rm BH}|^2 + |{\cal T}^{\rm DVCS}|^2 + {\cal I}
\, , \qquad
{\cal I}
= {\cal T}^{\rm DVCS} ( {\cal T}^{\rm BH} )^\ast
+ ( {\cal T}^{\rm DVCS} )^\ast {\cal T}^{\rm BH}
\, ,
\end{equation}
the Bethe-Heitler contribution $|{\cal T}^{\rm BH}|^2$ is merely an undesirable contamination
which was computed exactly already in Ref.\ \cite{BelMueKir01} and can be subtracted from the
cross section making use of the available vast data on the nucleon electromagnetic form factors
measured at facilities around the world. The main observables for extraction of GPDs emerge from
the remaining two contributions involving ${\cal T}^{\rm DVCS} $, the square of the DVCS amplitude
and the interference term ${\cal I}$.

In analogy to the hadronic electromagnetic current (\ref{EMcurrentFFs}), decomposed in terms of
the Dirac bilinears accompanied by the form factors, we parameterize the DVCS amplitude as follows
\begin{eqnarray}
\label{HadronicTensor-BMK}
T_{\mu\nu}
=
- {\cal P}_{\mu\sigma} g_{\sigma\tau} {\cal P}_{\tau\nu} \frac{q \cdot V_1}{p \cdot q}
+
\left( {\cal P}_{\mu\sigma} p_\sigma  {\cal P}_{\rho\nu}
+
{\cal P}_{\mu\rho}  p_\sigma {\cal P}_{\sigma\nu} \right)
\frac{V_{2\, \rho}}{p \cdot q}
-
{\cal P}_{\mu\sigma} i \varepsilon_{\sigma \tau q \rho} {\cal P}_{\tau\nu}
\frac{A_{1\, \rho}}{p \cdot q}
\, ,
\nonumber\\
\end{eqnarray}
where we have kept all dynamical contributions up to twist-three accuracy and, at the same time,
kinematically restored the electromagnetic gauge invariance exactly. Note, however, that the
so-called gluon transversity contribution, inducing the photon helicity-flip amplitude by two
units at leading twist level but suppressed by a power of $\alpha_s$, is not included here.  The
average four-momenta which enter this equation are $p = p_1 + p_2$ and $q = \ft12(q_1 + q_2)$.
The parametrization (\ref{HadronicTensor-BMK}) is similar to the one used in deep-inelastic
scattering. Indeed, the twist-two part of the generalized functions $V_1$ and $A_1$ corresponds
to the conventional $F_1$ and $g_1$ structure functions. The current conservation is ensured by
means of the projection operator
\begin{equation}
{\cal P}_{\mu\nu} = g_{\mu\nu} - \frac{q_{1 \mu} q_{2 \nu}}{q_1 \cdot q_2}
\, ,
\end{equation}
whose particular form is driven by the explicit calculation of the Compton amplitude via the
operator product expansion to twist-three accuracy \cite{PenPolShuStr00,BelMue00}, (see also Refs.\
\cite{AniPirTer00,RadWei00} for spinless targets). The $V_{2 \, \rho}$ structure is not independent
and is expressed in terms of the other two vector functions $V_{1 \, \rho}$ and $A_{1 \, \rho}$,
\begin{eqnarray}
\label{V2}
V_{2 \, \rho} = \xi
\left(
V_{1 \, \rho}
-
\frac{p_\rho}{2}
\frac{q \cdot V_1}{p \cdot q}
\right)
+
\frac{i}{2}
\frac{\varepsilon_{\rho\sigma\Delta q}}{p \cdot q} A_{1 \, \sigma}
\, ,
\end{eqnarray}
where $\xi = - q^2/p \cdot q$. The amplitudes $V_1$ and $A_1$ depend on the scaling variable $\Bx$,
the momentum transfer $\Delta^2$, and the hard momentum of the probe ${\cal Q}^2$, however, in order
to simplify our notations, we will drop this dependence when it is not essential for the presentation.
Their general decomposition in a complete basis of Compton form factors (CFFs) reads
\begin{eqnarray*}
\label{Def-V1}
V_{1\, \rho}\!\!\!&=&\!\!\!
\frac{1}{p \cdot q}
\bar{u}_2
\left(
{\not\!q}
\left[
p_\rho  {\cal H} + \Delta_{\perp \rho} {\cal H}^3_+
\right]
+
i \sigma_{\mu\nu} \frac{q_\mu \Delta_\nu}{2 M}
\left[
p_\rho  {\cal E}
+
\Delta_{\perp \rho} {\cal E}^3_+
\right]
+
\widetilde{\Delta}_{\perp\rho}
\left[
{\not\!q} \widetilde{{\cal H}}^3_-
+
\frac{q \cdot \Delta}{2 M}  \widetilde{\cal E}^3_-
\right]
\gamma_5
\right) u_1
\, , \\
\label{Def-A1}
A_{1\, \rho}\!\!\!&=&\!\!\!
\frac{1}{p \cdot q}
\bar{u}_2
\left(
{\not\!q}  \gamma_5
\left[
p_\rho  \widetilde{\cal H} + \Delta_{\perp \rho}  \widetilde{\cal H}^3_+
\right]
+
\frac{q \cdot \Delta}{2 M}
\gamma_5
\left[
p_\rho  \widetilde{\cal E}
+
\Delta_{\perp \rho} \widetilde{\cal E}^3_+
\right]
+
\widetilde\Delta_{\perp\rho}
\left[
{\not\!q}{{\cal H}}^3_-
+
i \sigma_{\mu\nu} \frac{q_\mu \Delta_\nu}{2 M}  {\cal E}^3_-
\right]
\right) u_1
\, ,
\end{eqnarray*}
again to twist-three accuracy. Here the CFFs  given by convolutions of perturbatively calculable
coefficient functions and a set of twist-two and -three GPDs (see Ref.\ \cite{BelMueKir01} for
details). In the above equations we use the following notations for the transverse components of
the $t$-channel momentum
$$
\Delta_\rho^\perp \equiv \Delta_\rho - \frac{\Delta\cdot q}{p \cdot q} p_\rho
\quad\mbox{and}\quad
\widetilde\Delta_\rho^\perp \equiv \frac{i\, \varepsilon_{\rho\Delta p q}}{p \cdot q}
$$
and where $\Delta\cdot q/p\cdot q \approx -\xi$ in DVCS kinematics.

\section{Helicity amplitudes}
\label{HelicityAmplitudes}

While in the BKM consideration \cite{BelMueKir01}, one is restricted to the twist-three
approximation for dynamical as well as kinematical effects, in the current analysis the
latter will be restored exactly since they account for the bulk of power-suppressed
corrections provided that there is a hierarchy of hadronic scales associated with higher-twist
operator matrix elements, such that, e.g., $\epsilon^2 \mbox{tw-2} \gg \ft{1}{Q^2} \mbox{tw-4}$.
An analysis of twist-four effects and higher is intrinsically involved due to complications and
ambiguities in the choice of operator bases. On the other hand, the incorporation of kinematical
power-suppressed effects is straightforward. In order to achieve this in the most efficient manner
we separate power corrections that arise from the leptonic and hadronic parts by evaluating photon
helicity amplitudes utilizing the polarization vectors for the incoming and outgoing photons in
the target rest frame. In addition to being a concise calculation scheme, it has an advantage of
localizing the azimuthal angle dependence in the lepton helicity amplitudes for the choice of the
reference frame with the $z$-axis counter-aligned with the incoming photon three-momentum. It also
allows for a straightforward reduction to the harmonic expansion introduced in Refs.\
\cite{BelMueKir01,DieGouPirRal97}.

We define the hadronic helicity amplitudes as
\be
\label{DVCS2helicity}
{\cal T}^{\rm DVCS}_{ac}(\phi) = (-1)^{a-1}
\varepsilon^{\mu\ast}_2(c) T_{\mu \nu} \varepsilon^{\nu}_1(a) \, ,
\ee
where the overall phase $(-1)^{a-1}$ accounts for the signature factor in the completeness relation for the
photon polarization vectors. These are constrained by the parity conservation and, as a consequence, we have
six independent functions,
\begin{eqnarray}
\label{helTsympro}
 {\cal T}^{\rm DVCS}_{--} ({\cal F})  &\!\!\!=\!\!\!&  {\cal T}^{\rm DVCS}_{++}({\cal F})
\Big|_{{\cal F}^{P=\pm 1}\to \pm {\cal F}^{P=\pm 1}}
\, , \nonumber\\
{\cal T}^{\rm DVCS}_{0-} ({\cal F})  &\!\!\!=\!\!\!&  {\cal T}^{\rm DVCS}_{0+}({\cal F})
\Big|_{{\cal F}^{P=\pm 1}\to \pm {\cal F}^{P=\pm 1}}
\, ,\\
 {\cal T}^{\rm DVCS}_{-+}({\cal F})  &\!\!\!=\!\!\!&   {\cal T}^{\rm DVCS}_{+-}({\cal F})
\Big|_{{\cal F}^{P=\pm 1}\to \pm {\cal F}^{P=\pm 1}}
\, . \nonumber
\end{eqnarray}
Substitution of the explicit parametrization for the Compton amplitude (\ref{HadronicTensor-BMK}) yields
dynamical twist-three approximation for the helicity amplitudes
\begin{eqnarray}
\label{cal-Tpplo}
{\cal T}^{\rm DVCS}_{aa}
\!\!\!&=&\!\!\!
\frac{1}{p \cdot q}
\left[ q \cdot V ({\cal F}) - a\, q \cdot A ({\cal F}) \right]
+ \mathcal{O} ({\cQ}^{- 2})\,,
\\
\label{cal-T0plo}
{\cal T}^{\rm DVCS}_{0a}
\!\!\!&=&\!\!\!
\frac{\sqrt{2}\widetilde{K}}{\cQ (2-\Bx)}
\frac{1}{p \cdot q} \left[
q\cdot V({\cal F}_{\rm eff})- a\, q\cdot A({\cal F}_{\rm eff})
\right]
+ \mathcal{O} ({\cQ}^{- 3}, \alpha_s {\cQ}^{- 1})
\, ,
\end{eqnarray}
where $a=\pm 1$ labels the helicity states of the final photon, ${\cal F}_{\rm eff}$
denotes the effective twist-three contribution in the notation of Ref.~\cite{BelMueKir01},
see Eqs.~(84)%
\footnote{We like to thank M.~Diehl for pointing out that the general relation (84) is
not applicable for the CFF ${\cal E}_{\rm eff}$. For this specific case we refer the reader
to our original work \cite{BelMue00}.}--(87) there, and
\begin{eqnarray}
{\widetilde K}
=
\sqrt{t_{\rm min} - t} \sqrt{(1 - \Bx)\sqrt{1 + \epsilon ^2}
+
\frac{(t_{\rm min} - t) \left(\epsilon^2 + 4 (1 - \Bx) \Bx \right)}{4 \cQ^2 }}
\, .
\end{eqnarray}
Note that the helicity flip amplitude ${\cal T}^{\rm DVCS}_{-+}$ arises from twist-two gluon
transversity, formally suppressed by $\alpha_s$, and higher twist contributions. Both of them
will not be considered here. In the following two sections we address the square of the DVCS
amplitude and the interference term in turn.

\subsection{Squared DVCS term}

Using the completeness relations for the photon polarization vectors, we can rewrite the square
of the DVCS amplitude
\begin{eqnarray}
|{\cal T}^{\rm DVCS}|^2
=
\frac{1}{{\cal Q}^2}
\sum_{a = {\scriptscriptstyle -}, 0, {\scriptscriptstyle +}}
\sum_{b = {\scriptscriptstyle -}, 0, {\scriptscriptstyle +}}
{\cal L}_{ab} (\lambda, \phi) {\cal W}_{ab}
\, , \quad
\end{eqnarray}
in terms of the hadronic,
\be
{\cal W}_{ab}
=
{\cal T}^{\rm DVCS}_{a +} \left({\cal T}^{\rm DVCS}_{b +}\right)^\ast
+
{\cal T}^{\rm DVCS}_{a -} \left({\cal T}^{\rm DVCS}_{b -}\right)^\ast
\,  ,
\ee
and leptonic,
\be
{\cal L}_{ab}(\lambda,\phi) =
\varepsilon^{\mu\ast}_1(a){\cal L}_{\mu \nu}(\lambda)\varepsilon^{\nu}_1(b)\,,
\ee
amplitudes, labeled by the helicity states of the initial photon. The latter
can be calculated exactly with the result already presented in Ref.\ \cite{BelMul09}
\begin{eqnarray}
{\cal L}_{++} (\lambda)
\!\!\!&=&\!\!\!
\frac{1}{ y^2(1 + \epsilon^2)} \left(2 - 2 y + y^2 + \frac{\epsilon ^2}{2} y^2 \right)
-
\frac{2-y}{\sqrt{1+\epsilon^2} y} \lambda
\, , \\
{\cal L}_{00}
\!\!\!&=&\!\!\!
\frac{4}{y^2(1+\epsilon^2)} \left(1-y - \frac{\epsilon ^2}{4} y^2\right)
\, , \\
{\cal L}_{0+}(\lambda,\phi)
\!\!\!&=&\!\!\!
\frac{2 - y - \lambda y \sqrt{1 + \epsilon^2}}{y^2 (1 + \epsilon^2)}
\sqrt{2} \sqrt{1 - y - \frac{\epsilon ^2}{4} y^2} \, e^{-i \phi}
\, , \\
{\cal L}_{-+}(\phi)
\!\!\!&=&\!\!\!
\frac{2}{y^2 (1 + \epsilon^2)} \left( 1-y - \frac{\epsilon ^2}{4} y^2 \right) e^{-i 2 \phi}
\,,
\end{eqnarray}
where the remaining amplitudes are related to the above ones by parity and time-reversal invariance,
\begin{eqnarray}
\begin{array}{ll}
{\cal L}_{0 -}(\lambda,\phi)
=
{\cal L}_{0+}(-\lambda,-\phi) \, ,
&
{\cal L}_{\pm,0}(\lambda,\phi)
=
{\cal L}_{0,\pm}(-\lambda,\phi)
\, , \qquad
\\[1mm]
{\cal L}_{--}(\lambda)
=
{\cal L}_{++}(-\lambda) \, ,
&
{\cal L}_{-+} (\phi)
=
{\cal L}_{+-}(-\phi)
\, .
\end{array}
\end{eqnarray}
More explicitly, neglecting transverse photon helicity-flip contributions, one finds for the
squared DVCS amplitude
\begin{eqnarray}
\label{HelAmpDVCS}
{\cal Q}^2|{\cal T}^{\rm DVCS}|^2 \!\!\!&=&\!\!\!
{\cal L}_{++}(\lambda)
{\cal T}^{\rm DVCS}_{++}  \left({\cal T}^{\rm DVCS}_{+ +}\right)^\ast
+{\cal L}_{++}(-\lambda)
{\cal T}^{\rm DVCS}_{--}  \left({\cal T}^{\rm DVCS}_{--}\right)^\ast
\nonumber\\
&+&\!\!\!{\cal L}_{00}\, \left[
{\cal T}^{\rm DVCS}_{0+}  \left({\cal T}^{\rm DVCS}_{0+}\right)^\ast
+
{\cal T}^{\rm DVCS}_{0-}  \left({\cal T}^{\rm DVCS}_{0-}\right)^\ast
\right]
\nonumber\\
&+&\!\!\!
{\cal L}_{0+}(\lambda,\phi)  {\cal T}^{\rm DVCS}_{0+}
\left({\cal T}^{\rm DVCS}_{+ +}\right)^\ast
+ {\cal L}_{0+}(-\lambda,-\phi) {\cal T}^{\rm DVCS}_{0-}
\left({\cal T}^{\rm DVCS}_{- -}\right)^\ast
\nonumber\\
&+&\!\!\!
{\cal L}_{0+}(\lambda,-\phi)
{\cal T}^{\rm DVCS}_{+ +}
\left({\cal T}^{\rm DVCS}_{0 +}\right)^\ast
+ {\cal L}_{0+}(-\lambda,\phi)
{\cal T}^{\rm DVCS}_{- -}
\left({\cal T}^{\rm DVCS}_{0 -}\right)^\ast
\, .
\end{eqnarray}

These findings immediately allow one to get the Fourier coefficients in the refined
approximation.  In addition to the overall prefactors
\begin{eqnarray}
\frac{1}{1+\epsilon^2} \qquad  \mbox{and}  \qquad   \frac{\lambda}{\sqrt{1+\epsilon^2}}\,,
\end{eqnarray}
accompanying the lepton helicity independent and dependent parts of the amplitude, respectively,
one find that the following substitutions in the refined approximation for the lepton-photon
``splitting kernels'',
\begin{eqnarray}
\label{Sub-DVCS-0-imp}
2-2 y + y^2 &\to&
2 - 2 y + y^2 + \ft12 \epsilon^2 y^2
\, , \\
1-y &\to&  1-y- \ft14 \epsilon^2 y^2
\, .
\nonumber
\end{eqnarray}
From here, we can read off the kinematically improved DVCS
harmonics in the decomposition
\begin{equation}
\label{AmplitudesSquared}
 |{\cal T}^{\rm DVCS}|^2
= \frac{e^6}{y^2 {\cal Q}^2}\left\{ c^{\rm DVCS}_0 + \sum_{n=1}^2
\left[ c^{\rm DVCS}_n \cos (n\phi) + s^{\rm DVCS}_n \sin (n \phi)
\right] \right\} \, ,
\end{equation}
with
\begin{eqnarray}
\label{Res-Mom-DVCS-0-imp} c_{0,\rm unp}^{\rm DVCS}
\!\!\!&=&\!\!\! 2 \frac{2-2 y + y^2+\frac{\epsilon ^2}{2}
y^2}{1+\epsilon^2} {\cal C}_{\rm unp}^{\rm DVCS} ({\cal F},{\cal
F}^\ast) + \frac{16 K^2}{(2-\Bx)^2 (1+\epsilon^2)} {\cal C}_{\rm
unp}^{\rm DVCS}({\cal F}_{\rm eff}, {\cal F}_{{\rm eff}}^\ast) \, ,
\nonumber\\ \\
\label{Res-Mom-DVCS-1-imp} \left\{ {c^{\rm DVCS}_{1,\rm unp} \atop
s^{\rm DVCS}_{1,\rm unp}} \right\} \!\!\!&=&\!\!\! \frac{8 K}{{(2
- \Bx)} (1+\epsilon^2)} \left\{ { (2 - y)  \atop - \lambda y
\sqrt{1 + \epsilon^2}} \right\} \left\{ {\Re{\rm e} \atop \Im{\rm
m}} \right\} \, {\cal C}^{\rm DVCS}_{\rm unp} \left( {\cal F}_{\rm
eff},{\cal F}^\ast\right)
\,
\end{eqnarray}
for an unpolarized target and
\begin{eqnarray}
\label{Res-Mom-DVCS-0-impLP} c_{0,{\rm LP}}^{\rm DVCS}
\!\!\!&=&\!\!\! \frac{2 \lambda \Lambda y
(2-y)}{\sqrt{1+\epsilon^2}} {\cal C}^{\rm DVCS}_{\rm LP} ({\cal F},{\cal F}^\ast)
\, ,
\\ \!\!\!& &\!\!\! \nonumber \\
\label{Res-Mom-DVCS-1-impLP} \left\{ { c^{\rm DVCS}_{1,{\rm LP}}
\atop s^{\rm DVCS}_{1,{\rm LP}} } \right\} \!\!\!&=&\!\!\!
-\frac{8 \Lambda K}{(2 - \Bx) (1+\epsilon^2)} \left\{{ - \lambda y
\sqrt{1 + \epsilon^2}  \atop  (2 - y) } \right\} \left\{ {\Re{\rm
e} \atop \Im{\rm m}} \right\} \, {\cal C}^{\rm DVCS}_{\rm LP}
\left( {\cal F}_{\rm eff},{\cal F}^\ast\right)
\end{eqnarray}
for the longitudinal polarized part, proportional to the polarization $\Lambda$. As in
Ref.~\cite{BelMueKir01}, we use the shorthand $$K = \sqrt{1-y + \frac{\epsilon ^2}{4}
y^2}\frac{{\widetilde K}}{\cQ}.$$ We emphasis that the squared twist-three contribution
in Eq.~(\ref{Res-Mom-DVCS-0-imp}) is a $1/\cQ^2$ suppressed contribution and that the
transversity contribution ${\cal F}_{T}$ is set to zero.

To evaluate the bilinear combinations ${\cal C}^{\rm DVCS}$ of CFFs, we rely on the
approximations (\ref{cal-Tpplo})--(\ref{cal-T0plo}). By means of Eq.~(\ref{HelAmpDVCS}),
we find the following result for the unpolarized
\begin{eqnarray}
\label{Def-CDVCSunp}
{\cal C}_{\rm unp}^{\rm DVCS} &\!\!\!=\!\!\!& \frac{\cQ^2
(\cQ^2+\Bx t)}{\left((2-\Bx) \cQ^2+ \Bx t\right)^2 } \Bigg\{
4(1-\Bx) {\cal H} {\cal H}^\ast + 4 \left(1-\Bx+\frac{2
\cQ^2+t}{\cQ^2+\Bx t}\frac{\epsilon ^2}{4} \right) \widetilde{\cal
H} \widetilde{\cal H}^\ast
\nonumber\\
&& -\frac{\Bx^2 (\cQ^2+t)^2}{\cQ^2 (\cQ^2+\Bx t)} \left( {\cal H}
{\cal E}^\ast + {\cal E} {\cal H}^\ast \right) -\frac{\Bx^2
\cQ^2}{\cQ^2+\Bx t} \left( \widetilde{\cal H} \widetilde{\cal
E}^\ast + \widetilde{\cal E} \widetilde{\cal H}^\ast
\right)\\
&& - \left( \frac{\Bx^2 \left(\cQ^2+t\right)^2 }{\cQ^2 (\cQ^2+\Bx
t)}+ \frac{\left((2-\Bx) \cQ^2+ \Bx t\right)^2 }{\cQ^2 (\cQ^2+\Bx
t)} \frac{t}{4 M^2} \right){\cal E} {\cal E}^\ast -\frac{\Bx^2
\cQ^2}{\cQ^2+\Bx t} \frac{t}{4 M^2} \widetilde{\cal E}
\widetilde{\cal E}^\ast \Bigg\} \nonumber
\end{eqnarray}
and longitudinally polarized combinations of CFFs
\begin{eqnarray}
\label{Def-CDVCSLP}
{\cal C}_{\rm LP}^{\rm DVCS} &\!\!\!=\!\!\!& \frac{\cQ^2(\cQ^2+\Bx
t)}{ \sqrt{1+\epsilon^2} \left((2-\Bx) \cQ^2+\Bx t\right)^2}
\Bigg\{ 4\!\left(\!1-\Bx+ \frac{(3-2 \Bx) \cQ^2+ t}{\cQ^2+\Bx t}
\frac{\epsilon^2}{4} \right)\!\left( {\cal H} \widetilde{\cal H}^\ast +
\widetilde{\cal H} {\cal H}^\ast \right)
\nonumber\\
&& - \frac{\cQ^2-\Bx (1-2 \Bx) t}{\cQ^2+\Bx t }\, \Bx^2 \left( {\cal
H} \widetilde{\cal E}^\ast + \widetilde{\cal E} {\cal H}^\ast +
\widetilde{\cal H} {\cal E}^\ast + {\cal E} \widetilde{\cal
H}^\ast
\right)\\
&& -\frac{4 (1-\Bx) \left(\cQ^2+ \Bx t\right) t   +
\left(\cQ^2+t\right)^2 \epsilon^2 }{2 \cQ^2 \left(\cQ^2+ \Bx
t\right)}\, \Bx \left( \widetilde{\cal H} {\cal E}^\ast + {\cal E}
\widetilde{\cal H}^\ast \right)
\nonumber\\
&& - \frac{(2-\Bx) \cQ^2+\Bx t}{\cQ^2+\Bx t}
\left(\frac{\Bx^2\left(\cQ^2+t\right)^2 }{2\cQ^2 \left((2-\Bx)
\cQ^2+ \Bx t\right)}+\frac{t}{4 M^2}\right)\,\Bx  \left( {\cal E}
\widetilde{\cal E}^\ast + \widetilde{\cal E} {\cal E}^\ast \right)\!
\Bigg\} \, ,
\nonumber
\end{eqnarray}
respectively. The uncertainties from remaining kinematical and dynamical higher twist
contributions are included in the bilinear combinations ${\cal C}^{\rm DVCS}$ of CFFs.
As shown in Ref.~\cite{BelMue08} for a (pseudo) scalar target, i.e., setting
$$
\frac{q\cdot V({\cal F})}{ q \cdot p} = {\cal H}
\, , \quad
\frac{q\cdot V({\cal F}_{\rm eff})}{ q \cdot p} = {\cal H}_{\rm eff}
\, , \quad q \cdot A = 0
\, ,
$$
in Eqs.~(\ref{cal-Tpplo})--(\ref{cal-T0plo}), different parameterizations of the DVCS
amplitude result only in small numerical deviations even at rather low energy and photon
virtualities. Finally, neglecting $1/\cQ^2$ power suppressed terms in the presented
findings for the squared DVCS amplitude leads to those of Ref.~\cite{BelMueKir01}.

\subsection{Interference term}

Let us now treat the interference term  in a manner completely analogous to the consideration
of the squared DVCS amplitude given above. Inserting the completeness condition for the initial
and final photon polarization states, one finds ${\cal I}$ as a linear superposition
\begin{eqnarray}
\label{Def-Sqa-Int-Hel}
{\cal I} =  \frac{\pm e^6}{t \, {\cal P}_1(\phi) {\cal P}_2(\phi)}
\sum_{a = {\scriptscriptstyle -}, 0, {\scriptscriptstyle +}}
\sum_{b = {\scriptscriptstyle -}, {\scriptscriptstyle +}}
\sum_{S^\prime}
\left\{
{\cal L}^\rho_{a b}(\lambda,\phi)  {\cal T}_{a b}J^\dagger_\rho
+
\left({\cal L}^\rho_{a b}(\lambda,\phi) {\cal T}_{a b}J^\dagger_\rho\right)^\ast \right\}
\, ,
\end{eqnarray}
of products of hadronic and leptonic helicity amplitudes. The former were defined earlier in Eqs.\
(\ref{cal-Tpplo})--(\ref{cal-T0plo}) while the latter read
\begin{eqnarray}
{\cal L}^\rho_{a b}(\lambda,\phi)
=
\varepsilon_1^{\mu\ast}(a) L_{\mu\phantom{\rho}\nu}^{\phantom{\mu}\rho}  \varepsilon_2^\nu(b)
\, .
\end{eqnarray}
Summation over the final nucleon polarization states yields the following result for the building blocks
of the hadronic amplitudes (\ref{cal-Tpplo})--(\ref{cal-T0plo})
\begin{eqnarray}
\label{Def-Intsum-unp}
\sum_{S^\prime}    \frac{q \cdot V}{p \cdot q} J^\dagger_\rho
\!\!\!&=&\!\!\!
p_{\rho} \left[ {\cal C}^{{\cal I}}_{\rm unp} - {\cal C}^{{\cal I},A}_{\rm unp}  \right]\! ({\cal F})
+
2 q_{\rho}  \frac{t}{\cQ^2}  {\cal C}^{{\cal I},V}_{\rm unp} ({\cal F})
+
\frac{2 \Lambda}{\sqrt{1+\epsilon^2}}   \frac{i \varepsilon_{pq\Delta\rho}}{\cQ^2} \,
{\cal C}^{{\cal I},V}_{\rm LP} ({\cal F})  \,,
\\
\label{Def-Intsum-LP}
\sum_{S^\prime}  \frac{q \cdot A}{p \cdot q} J^\dagger_\rho
\!\!\!&=&\!\!\!
\frac{\Lambda  p_{\rho}}{\sqrt{1+\epsilon^2}}
\left[{\cal C}^{{\cal I}}_{\rm LP} - {\cal C}^{{\cal I},V}_{\rm LP}\right] \!({\cal F})
+
2 q_\rho \frac{\Lambda}{\sqrt{1+\epsilon^2}} \frac{t}{\cQ^2}\, {\cal C}^{{\cal I},A}_{\rm LP} ({\cal F})
+
2 \frac{i \varepsilon_{pq\Delta\rho}}{\cQ^2}\, {\cal C}_{\rm unp}^{{\cal I},A} {(\cal F})
\, .
\end{eqnarray}
Here, to match the notation of Ref.~\cite{BelMueKir01}, we introduced the following combination of CFFs
\begin{eqnarray}
\label{def-Cunp}
{\cal C}_{\rm unp}^{{\cal I}} {(\cal F})
\!\!\!&=&\!\!\!
F_1 {\cal H} - \frac{t}{4 M^2}  F_2 {\cal E}
+
\frac{\Bx}{2 - \Bx + \Bx \frac{t}{{\cQ}^2}}
(F_1+F_2)\widetilde{\cal H}  \,,
\\
\label{def-CunpDel}
{\cal C}^{{\cal I},V}_{\rm unp} ({\cal F})
\!\!\!&=&\!\!\!
\frac{\Bx}{2 - \Bx + \Bx \frac{t}{{\cQ}^2}}   (F_1+F_2) ( {\cal H}+  {\cal E})  \, ,
\\
 {\cal C}^{{\cal I},A}_{\rm unp} ({\cal F})
\!\!\! &=&\!\!\! \frac{\Bx}{2 - \Bx + \Bx \frac{t}{{\cQ}^2}}  (F_1+F_2)\widetilde{\cal H} \,,
\\
\label{def-CLP}
{\cal C}^{{\cal I}}_{\rm LP} ({\cal F})
\!\!\!&=&\!\!\!
\frac{\Bx}{2 - \Bx + \Bx \frac{t}{{\cQ}^2}}
(F_1+F_2)\left[{\cal H}+ \frac{\Bx}{2} \left(1-\frac{t}{\cQ^2}\right) {\cal E}\right]
\\
&+&\!\!\!
 \left[
1+ \frac{M^2}{\cQ^2} \frac{\Bx^2}{2 - \Bx + \Bx \frac{t}{{\cQ}^2}}  \left( 3 + \frac{t}{ \cQ^2} \right)
\right]  F_1 \widetilde{\cal H}
-
\frac{t}{\cQ^2}   \frac{2 \Bx (1 - 2 \Bx)}{2 - \Bx + \Bx \frac{t}{{\cQ}^2}} F_2  \widetilde{\cal H}
\nonumber\\
&-&\!\!\!
\frac{\Bx}{2 - \Bx + \Bx \frac{t}{{\cal Q}^2}}
\left[ \frac{\Bx}{2} \left( 1 - \frac{t}{\cQ^2} \right) F_1  +\frac{t}{4 M^2} F_2 \right] \widetilde{\cal E}\,,
\nonumber\\
\label{def-CLPDelV}
{\cal C}^{{\cal I},V}_{\rm LP} ({\cal F})
\!\!\!&=&\!\!\!
\frac{\Bx}{2 - \Bx + \Bx \frac{t}{{\cQ}^2}}
(F_1+F_2)\left[{\cal H}+ \frac{\Bx}{2} \left(1-\frac{t}{\cQ^2}\right) {\cal E}\right],
\\
\label{def-CLPDelA}
{\cal C}^{{\cal I},A}_{\rm LP} ({\cal F})
\!\!\!&=&\!\!\!
\frac{\Bx}{2 - \Bx + \Bx \frac{t}{{\cQ}^2}}  (F_1+F_2)
\left[ \widetilde{\cal H}  + 2\Bx \frac{M^2}{\cQ^2} \widetilde{\cal H}
+ \frac{\Bx}{2} \widetilde{\cal E} \right]
\, .
\end{eqnarray}
Note that the ambiguity in the parameterization of hadronic helicity amplitudes
(\ref{cal-Tpplo})--(\ref{cal-T0plo}) is also exhibited in the $q_{\rho}$-structure
of  Eqs.~(\ref{Def-Intsum-unp})--(\ref{Def-Intsum-LP}), which are kinematically
suppressed by $t/\cQ^2$.  Such terms appear in the azimuthal angle independent
part of the interference term at ``twist-three'' level, yielding the addenda
$$
\Delta{\cal C}^{\cal I}_{\rm unp}({\cal F})
=
- \lim_{{\cal Q}\to \infty} \left[
\frac{\Bx}{2-\Bx} {\cal C}_{\rm unp}^{{\cal I},V}  + {\cal C}_{\rm unp}^{{\cal I},A}\right]({\cal F})
\, , \quad
\Delta{\cal C}^{\cal I}_{\rm LP}({\cal F})
=
-\lim_{{\cal Q}\to \infty} \left[
 {\cal C}_{\rm LP}^{{\cal I},V} + \frac{\Bx}{2-\Bx} {\cal C}_{\rm LP}^{{\cal I},A}\right]({\cal F})
\, .
$$
As a cross check, neglecting the power suppressed contributions yields the CFF and FF
combinations that appear in ${\cal C}^{\cal I}({\cal F})$ and $\Delta{\cal C}^{\cal I}
({\cal F})$  of Ref.\ \cite{BelMueKir01}.

Now we turn to the leptonic helicity amplitudes which contain the entire azimuthal angular dependence of
the interference term. Their contraction with the hadronic amplitude with respect to the Lorentz indices
introduces the Fourier harmonics in the definition (\ref{Def-Sqa-Int-Hel}) of the interference term yields
\begin{equation}
\label{InterferenceTerm}
{\cal I}
= \frac{\pm e^6}{\Bx y^3 t {\cal P}_1 (\phi) {\cal P}_2 (\phi)}
\left\{
c_0^{\cal I}
+ \sum_{n = 1}^3
\left[
c_n^{\cal I} \cos(n \phi) +  s_n^{\cal I} \sin(n \phi)
\right]
\right\} \, ,
\end{equation}
with kinematically power-suppressed contributions exactly accounted for,
\begin{eqnarray}
\label{Res-IntTer-unp-c-im}
c^{\cal I}_{n}\!\!\!&=&\!\!\! C_{++} (n) \, \Re{\rm e} \, {\cal C}_{++}^{{\cal I}} \left(n|{\cal F}\right)
+
C_{0+} (n) \, \Re{\rm e} \, {\cal C}_{0+}^{{\cal I}}
\left(n| {\cal F}_{\rm eff} \right)
+
C_{-+} (n) \, \Re{\rm e} \, {\cal C}_{-+}^{{\cal I}} \left(n| {\cal F}_T \right)
\, , \nonumber\\
\\
\label{Res-IntTer-unp-s-im}
\!\!\!s^{\cal I}_{n}\!\!\!&=&\!\!\!
S_{++} (n) \, \Im{\rm m} \, {\cal S}^{{\cal I}}_{++} \left(n|{\cal F}\right)
+
S_{0+} (n) \, \Im{\rm m} \, {\cal S}^{{\cal I}}_{0+}
\left(n|{\cal F}_{\rm eff}\right)
+
S_{-+} (n) \, \Im{\rm m} \, {\cal S}_{-+}^{\cal I} \left(n|{\cal F}_T\right)
\, . \nonumber
\end{eqnarray}
The above coefficients are defined in terms of the photon helicity-conserving
\begin{eqnarray}
{\cal C}_{++}^{{\cal I}} \left(n|{\cal F}\right)  &\!\!\!= \!\!\!&
{\cal C}^{{\cal I}} \left({\cal F}\right) + \frac{C^V_{++} (n) }{ C_{++} (n) }
{\cal C}^{{\cal I},V} \left({\cal F}\right) +  \frac{C^A_{++} (n) }{ C_{++} (n) }
{\cal C}^{{\cal I},A} \left({\cal F}\right)
\nonumber\\
{\cal S}_{++}^{{\cal I}} \left(n|{\cal F}\right)  &\!\!\!= \!\!\!&
{\cal C}^{{\cal I}} \left({\cal F}\right) + \frac{S^V_{++} (n) }{ S_{++} (n) }
{\cal C}^{{\cal I},V} \left({\cal F}\right) +  \frac{S^A_{++} (n) }{ S_{++} (n) }
{\cal C}^{{\cal I},A} \left({\cal F}\right)
\end{eqnarray}
and helicity-changing amplitudes
\begin{eqnarray}
{\cal C}_{0+}^{{\cal I}} \left(n|{\cal F}_{\rm eff}\right)  &\!\!\!= \!\!\!&
\frac{\sqrt{2}}{2-\Bx}\, \frac{\widetilde{K}}{\cQ}\left[
{\cal C}^{{\cal I}} \left({\cal F}_{\rm eff}\right) + \frac{C^V_{0+} (n) }{ C_{0+} (n) }
{\cal C}^{{\cal I},V} \left({\cal F}_{\rm eff}\right) +  \frac{C^A_{0+} (n) }{ C_{0+} (n) }
{\cal C}^{{\cal I},A} \left({\cal F}_{\rm eff}\right)
\right]
\nonumber\\
{\cal S}_{0+}^{{\cal I}} \left(n|{\cal F}_{\rm eff}\right)  &\!\!\!= \!\!\!&
\frac{\sqrt{2}}{2-\Bx}\, \frac{\widetilde{K}}{\cQ}\left[
{\cal C}^{{\cal I}} \left({\cal F}_{\rm eff}\right) + \frac{S^V_{0+} (n) }{ S_{0+} (n) }
{\cal C}^{{\cal I},V} \left({\cal F}_{\rm eff}\right) +  \frac{S^A_{0+} (n) }{ S_{0+} (n) }
{\cal C}^{{\cal I},A} \left({\cal F}_{\rm eff}\right)
\right]
\, ,
\end{eqnarray}
respectively. For an unpolarized target the coefficients $C_{a b} (n) $ and $S_{a b} (n)$ were
already known from the study of a (pseudo) scalar target \cite{BelMue08}. The complete set of coefficients
$C^{\mathcal I}_{a b} (n)$ and $S^{\mathcal I}_{a b} (n)$ is given in Appendix \ref{FourierHarmonics}.

\section{Discussion and conclusions}

Let us shortly summarize our framework.  To separate leptonic and hadronic contributions,  we
defined helicity amplitudes in a specific reference frame that is commonly used to confront
experimental measurements and theoretical predictions.  Within this convention, the leptonic
part was calculated exactly. As far as the hadronic part is concerned, a few comments are in order.
\begin{itemize}
\item
To evaluate the hadronic part, we employed the parametrization (\ref{cal-Tpplo})--(\ref{cal-T0plo}).
The $1/\cQ^2$-suppressed terms in both the bilinear (\ref{Def-CDVCSunp})--(\ref{Def-CDVCSLP}) and
linear combinations (\ref{def-CunpDel})--(\ref{def-CLPDelA}) of CFFs  mainly arise from the exact
treatment of the hadronic states, including parameterization of the polarization vector.
\item There are intrinsic twist-four uncertainties in the above definitions, induced by the
parameterization of the light cone projection, i.e.,  $(n \cdot V)$ and $(n \cdot A)$ in terms
of the four-vectors defining the process kinematics and by missing pieces in the DVCS tensor that
are needed for the restoration of the electromagnetic current conservation, see discussion in
Ref.~\cite{BelMue08}.
\item Assuming that there is a hierarchy of hadronic scales, associated with higher-twist operator
matrix elements, we mainly kept power suppressed twist-two contributions while neglecting genuine
dynamical twist-four effects, i.e., $\ft{t}{Q^2} \mbox{tw-2} \gg \ft{1}{Q^2} \mbox{tw-4}$.
\end{itemize}

\begin{figure}[t]
\begin{center}
\mbox{
\begin{picture}(0,320)(250,0)
\put(20,170){\insertfig{7.65}{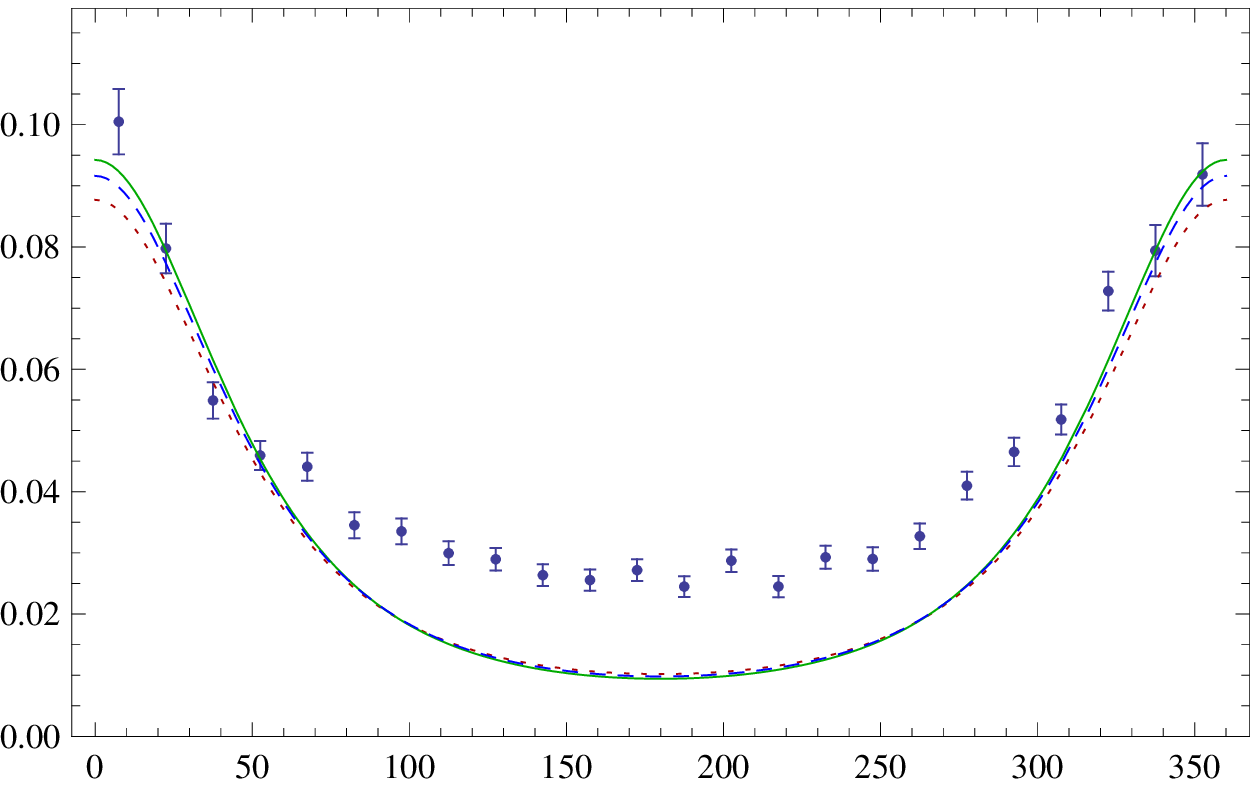}}
\put(0,205){\rotatebox{90}{$d^4 \sigma [{\rm nb}/{\rm GeV}^4]$}}
\put(206,160){$\phi$} \put(270,170){\insertfig{7.65}{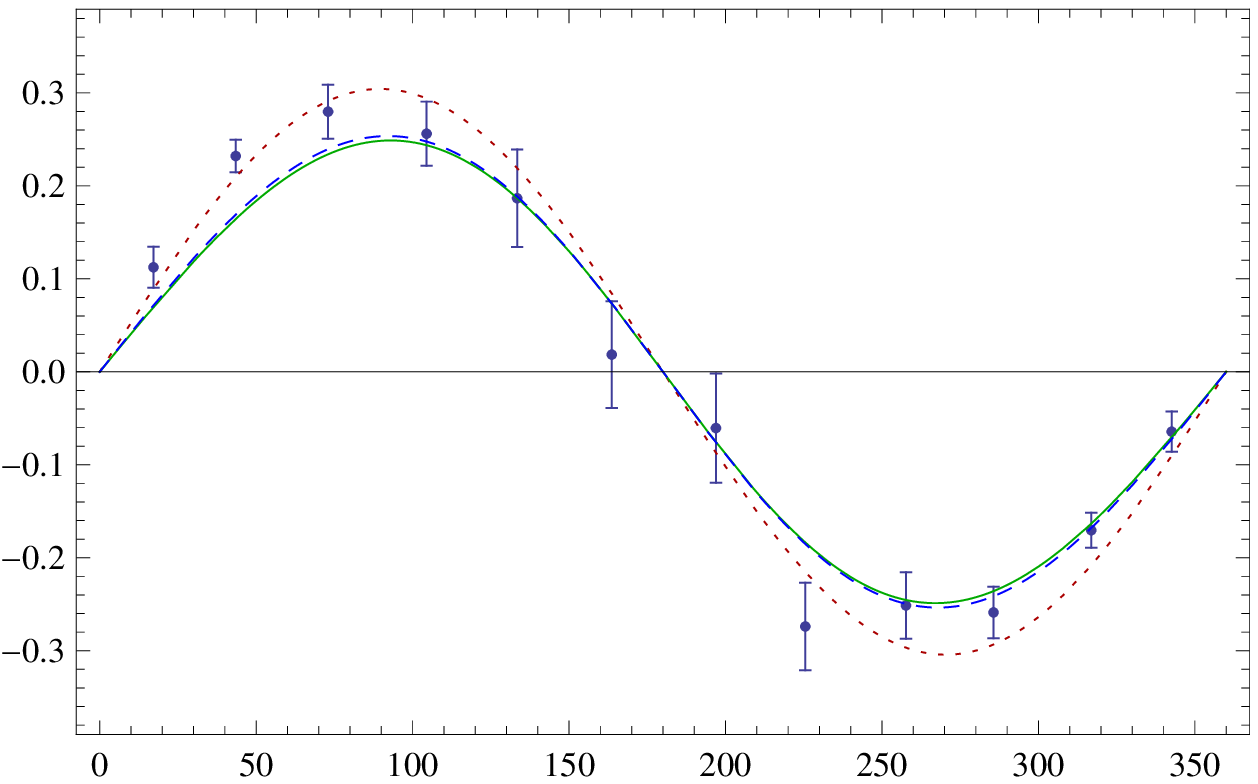}}
\put(255,228){\rotatebox{90}{BSA}} \put(458,160){$\phi$}
\put(24,0){\insertfig{7.50}{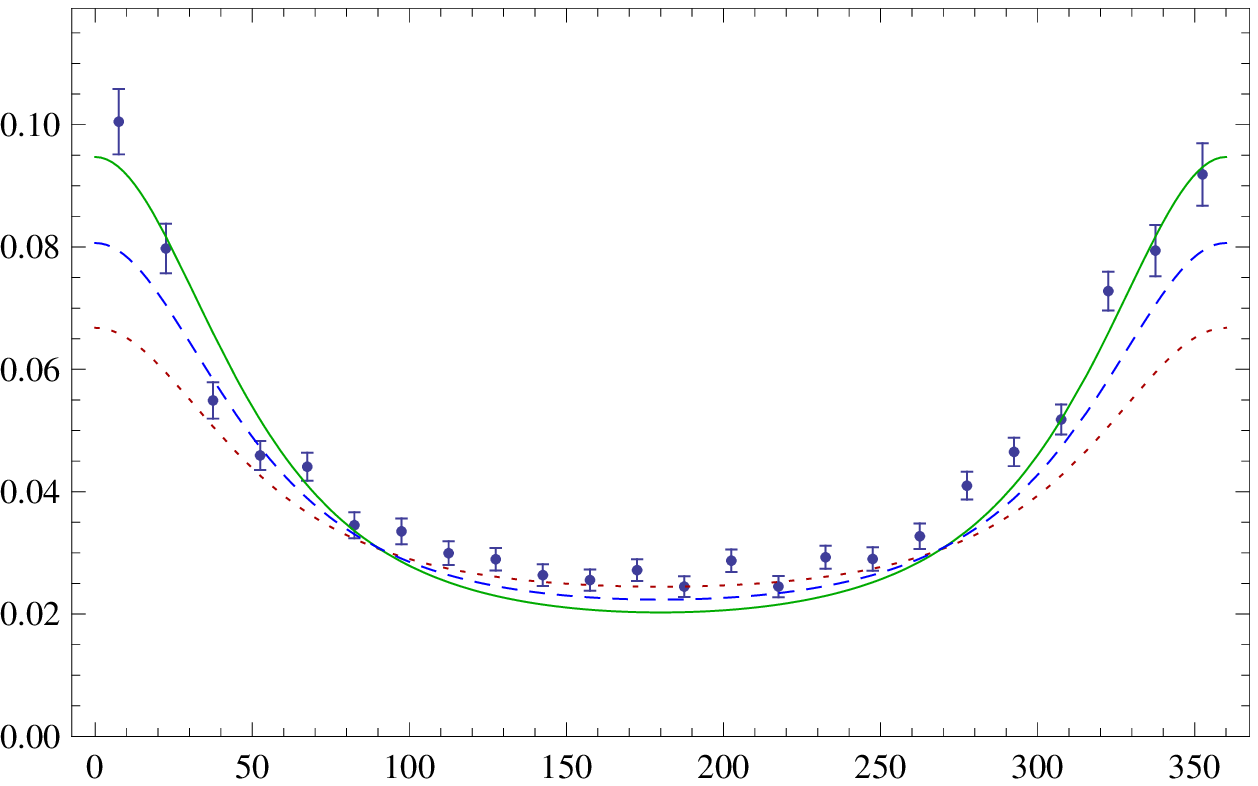}}
\put(0,35){\rotatebox{90}{$d^4 \sigma [{\rm nb}/{\rm GeV}^4]$}}
\put(206,-10){$\phi$} \put(270,0){\insertfig{7.65}{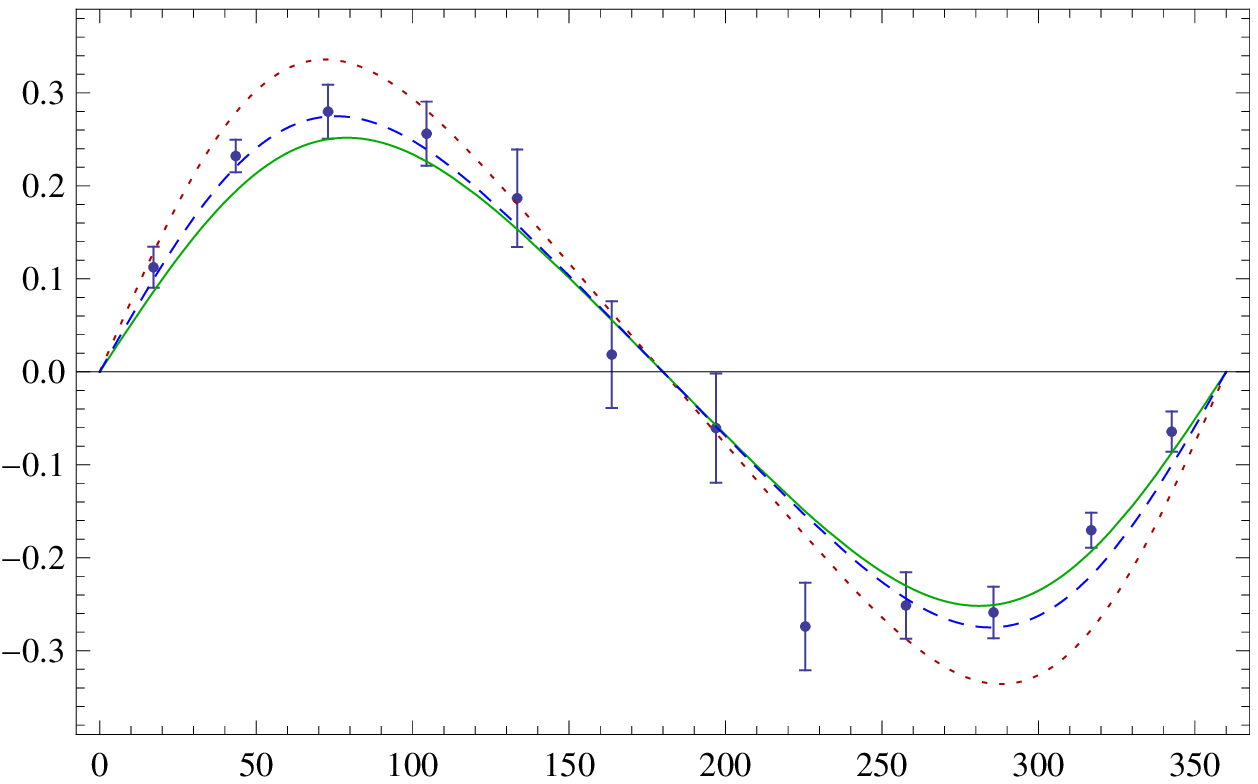}}
\put(255,58){\rotatebox{90}{BSA}} \put(458,-10){$\phi$}
\end{picture}
}
\end{center}
\caption{\label{Estimates} GPD predictions, resulting from two fits of Ref.\ \cite{KumMul09} as
described in the main body of the paper, for the BKM approximation (red dotted) to the exact
(green solid) and ``hot fix'' (blue dashed) for the four-fold cross section (\ref{WQ}) evaluated
for lepton beam of positive helicity $\lambda = 1$ in the left panels and the beam-spin asymmetry
(\ref{BSA}) in the right panels.
}
\end{figure}

Let us now explore the magnitude of power suppressed effects we have accounted for in this
work compared to the approximate treatment of the older analysis in Ref.\ \cite{BelMueKir01}.
Before we present our predictions, let us introduce a simplified treatment of exact kinematical
correction to the BKM formalism, which can be regarded as an improvement of the BKM analysis
for an unpolarized target. This consists of replacing the BKM coefficients, entering the angular
dependence of the cross section, with exact ones from the spin-zero case and ignoring at the
same time all other induced harmonics for the same hadronic helicity amplitudes. Moreover, the
BKM expressions for the hadronic ${\cal C}$-coefficients are taken. We will dub this scheme as
``hot fix''. It consists of substitutions
\be
c^{{\cal I}}_n |_{\rm BKM} \to
C_{++} (n) \, \Re{\rm e} \, {\cal C}_{++}^{{\cal I}} \left(n|{\cal F}\right)
\, , \qquad
s^{{\cal I}}_n |_{\rm BKM} \to
S_{++} (n) \, \Im{\rm m} \, {\cal S}^{{\cal I}}_{++} \left(n|{\cal F}\right)
\, ,
\ee
with expression on the right-hand sides given in Appendix \ref{FourierHarmonics}. It provides
a very accurate description of experimental observables in favorable situations. To demonstrate
our point, we evaluated the four-fold cross section (\ref{WQ}) and the beam-spin asymmetry
defined as
\be
\label{BSA}
{\rm BSA}
=
\frac{
d^4 \sigma (\lambda = +1) - d^4 \sigma (\lambda = - 1)
}{
d^4 \sigma (\lambda = +1) - d^4 \sigma (\lambda = - 1)
}
\, ,
\ee
where power suppressed corrections in the hadronic sector are still neglected.
For illustration, we show in Figure \ref{Estimates} predictions for approximation schemes advocated
in this paper for two dispersive approach fits \cite{KumMul09} to the available experimental data,
where twist-three and gluon transversity CFFs were projected out. (Thus, the observables, shown in
the figure, are not the ones used in the fit, rather they contain an admixture of higher harmonics.)
In the left panels, we display the unpolarized cross section measurement of Jefferson Laboratory's
Hall A \cite{HallA06} for the kinematics
$$
E = 5.75 \ {\rm GeV} \, , \quad {\cal Q}^2 = 2.3 \ {\rm GeV}^2 \,
, \quad t = - 0.36 \ {\rm GeV}^2 \, , \quad \Bx = 0.36 \, ,
$$
while on the right ones a beam spin measurement of the CLAS collaboration \cite{HallB07} is shown for
$$
E = 5.77 \ {\rm GeV} \, , \quad {\cal Q}^2 = 1.95 \ {\rm GeV}^2 \,
, \quad t = - 0.28 \ {\rm GeV}^2 \, , \quad \Bx = 0.25 \, .
$$
In the upper panels, the predictions are given for a fit that excluded the Hall A data \cite{HallA06}
and assumed the dominance of the unpolarized GPD $H$ in the DVCS amplitude. It is clearly demonstrated
that (with present understanding of GPD magnitude) such a hypothesis is in conflict with the data. In
the lower panels, the unpolarized cross section measurements of Hall A were included, with the fit
performed to the ratio of $\cos(1\cdot\phi)$ and $\cos(0\cdot \phi)$ harmonics of the weighted cross
section, see Eq.~(103) of Ref.~\cite{BelMueKir01}, rather than to the cross section itself. To describe
the data, one required a large real part in the DVCS amplitude which was {\sl effectively} obtained from
an abnormally large contribution of the GPD $\widetilde H$. As we demonstrate in the figure, in the
former case, the difference between the BKM (dotted) and exact (solid) results, while more accidentally
of order of a few percent in the cross section (except for the end-point regions), reaches $20-25\%$ in
the BSA asymmetry (right panel). However, the deviations of the ``hot fix'' (dashed) from the exact
treatment is vanishingly small. Confronting the two fits, done with different dynamical assumptions about
contributing GPDs, exhibits first of all larger effects of power suppressed correction in the differential
cross section in the lower compared to upper panels, and second, demonstrates significant differences
between the ``hot fix'' (dashed) and exact (solid) results. In other words, in a fitting procedure
relying on cross section formulas with exact treatment of kinematical effects rather than the ones based
on a hot fix, one anticipates that the magnitude of $\widetilde H$ becomes smaller.

The improvement on the BKM approximation scheme that we advocated in this paper demonstrate the necessity
to incorporate power-suppressed corrections stemming from the kinematical effects in the leptonic part of
the electroproduction scattering amplitudes. The results for gluon transversity  and transversal polarized
target will be presented somewhere else. The next set of problems of paramount importance is to develop
a calculational scheme for analysis of dynamical higher twist correlation functions contributing to the
DVCS amplitude, echoing formalism developed before for deep-inelastic scattering \cite{EllFurPet82,BalBra88}
as well as target mass and momentum transfer corrections, extending earlier result beyond the leading twist
order \cite{BelMul01,GeyRobEil05}.

\vspace{0.5cm}

\noindent

\appendix

\section{Fourier harmonics of the leptonic tensor}
\label{FourierHarmonics}

Let us present explicit expressions for the Fourier coefficients entering the leptonic part of the
interference term \re{Def-Sqa-Int-Hel}.

\subsection{Unpolarized target}
\label{FourierHarmonics-unp}

The angular coefficients  $C^{\rm unp}_{ab} (n)$ and $S^{\rm unp}_{ab} (n)$  for unpolarized target are
given by the expressions $C_{ab} (n)$ and $\lambda S_{ab} (n)$   for scalar target \cite{BelMue08}, while
the results for $C^{{\rm unp},V}_{ab} (n), C^{{\rm unp},A}_{ab} (n), S^{{\rm unp},V}_{ab} (n)$ and
$S^{{\rm unp},A}_{ab} (n)$ are new.  The third odd harmonics vanishes, i.e.,
$$
S^{\rm unp}_{ab} (n=3) = S^{{\rm unp},V}_{ab} (n=3) = S^{{\rm unp},A}_{ab} (n=3)=0 \, ,
$$
and will be not listed.

\noindent
\underline{Conserved photon-helicity coefficients:}
\begin{eqnarray}
C_{++}^{\rm unp} (n \!\!\!&=&\!\!\! 0)
=
-\frac{4(2 - y) \left(1+\sqrt{1+\epsilon^2}\right)}{(1+\epsilon^2)^{2}}
\Bigg\{
\frac{{\widetilde K}^2}{\cQ^2} \frac{(2-y)^2 }{\sqrt{1+\epsilon^2}}
\\
&+&\!\!\!
\frac{t}{\cQ^2} \left(1-y-\frac{\epsilon^2}{4} y^2 \right)(2-\Bx)
\Bigg(
1
+
\frac{2\Bx \left(2-\Bx + \frac{\sqrt{1+\epsilon^2}-1}{2}+ \frac{\epsilon^2}{2\Bx}\right)\frac{t}{\cQ^2}
+
\epsilon^2}{(2-\Bx)(1+\sqrt{1+\epsilon^2})}
\Bigg)
\Bigg\}
\, ,
\nonumber\\
C^{{\rm unp},V}_{++} (n \!\!\!&=&\!\!\! 0)=
\frac{8 (2-y) }{(1+\epsilon ^2)^2}\frac{\Bx t}{\cQ^2}
\Bigg\{
\frac{(2-y)^2 {\widetilde K}^2}{\sqrt{1+\epsilon^2}\cQ^2 } +
\left(1-y-\frac{\epsilon^2}{4} y^2 \right)\frac{1+\sqrt{1+\epsilon ^2}}{2}
\nonumber\\
& &\hspace{5cm}\times \left(1+\frac{t}{\cQ^2}\right)  \left(1+
\frac{\sqrt{1+\epsilon ^2}-1+2 \Bx}{1+\sqrt{1+\epsilon^2}} \frac{t}{\cQ^2} \right)
 \Bigg\},
\nonumber\\
 C^{{\rm unp},A}_{++} (n \!\!\!&=&\!\!\! 0)=
\frac{8(2-y)}{(1+\epsilon^2)^2} \frac{t}{\cQ^2}
\Bigg\{
\frac{(2-y)^2 {\widetilde K}^2}{\sqrt{1+\epsilon^2}\cQ^2}\frac{1+\sqrt{1+\epsilon^2} -2\Bx}{2}
 +
 \left(1-y-\frac{\epsilon^2}{4} y^2 \right)
\Bigg[ \frac{1+\sqrt{1+\epsilon^2} }{2}
\nonumber\\
&&\times
\left(
1+\sqrt{1+\epsilon^2}-\Bx +
\left(
\sqrt{1+\epsilon^2}-1+\Bx \frac{3+\sqrt{1+\epsilon^2} -2\Bx}{1+\sqrt{1+\epsilon^2} }
\right)\frac{t}{\cQ^2}
\right)- \frac{2{\widetilde K}^2}{\cQ^2}\Bigg]
\Bigg\},
\nonumber\\
\phantom{\frac{a}{b}}
\nonumber\\
C_{++}^{\rm unp} (n \!\!\!&=&\!\!\! 1)
=
\frac{-16 K\left(1-y-\frac{\epsilon^2}{4} y^2 \right)}{(1+\epsilon^2)^{5/2}}
\left\{
\left(
1+ (1-\Bx)  \frac{\sqrt{\epsilon ^2+1}-1}{2 \Bx} + \frac{\epsilon ^2}{4 \Bx}
\right)
\frac{\Bx t}{\cQ^2} - \frac{3 \epsilon ^2}{4}
\right\}
\nonumber\\
&-&\!\!\!\!\!
4 K \left(2-2 y+y^2+ \frac{\epsilon^2}{2}y^2\right)
\frac{1+\sqrt{1+\epsilon ^2}-\epsilon ^2}{(1+\epsilon^2)^{5/2}}
\Bigg\{\!
1 - (1-3\Bx) \frac{t}{{\cal Q}^2}
\nonumber\\
&&\qquad\qquad\qquad\qquad\qquad\qquad\qquad\qquad\qquad\qquad
+
\frac{1 - \sqrt{1 + \epsilon^2} + 3 \epsilon^2}{1 + \sqrt{1 + \epsilon^2} - \epsilon^2} \frac{\Bx t}{\cQ^2}
\Bigg\},
\nonumber \\
 C^{ {\rm unp},V}_{++} (n \!\!\!&=&\!\!\! 1)=
\frac{16 K}{(1+\epsilon ^2)^{5/2}} \frac{ \Bx t}{\cQ^2}
\Bigg\{
(2-y)^2 \left(1-(1-2\Bx) \frac{t}{\cQ^2} \right)+
\left(1-y-\frac{\epsilon^2}{4} y^2 \right)
\nonumber\\
&&\hspace{5cm}\times
\frac{1+\sqrt{1+\epsilon ^2}-2 \Bx}{2}\frac{t^\prime}{\cQ^2}
\Bigg\},
\nonumber\\
 C^{{\rm unp},A}_{++} (n \!\!\!&=&\!\!\! 1)=
\frac{-16 K }{(1+\epsilon^2)^2} \frac{t}{\cQ^2}
\Bigg\{
\left(1-y-\frac{\epsilon^2}{4} y^2 \right)\left(1-(1-2\Bx)\frac{t}{\cQ^2}+\frac{4\Bx(1-\Bx)+\epsilon^2}{4\sqrt{1+\epsilon^2}}\frac{t^\prime}{\cQ^2}\right)
\nonumber\\
&&\quad
-(2-y)^2
\left(
1-\frac{\Bx}{2} +\frac{1+\sqrt{1+\epsilon^2}-2\Bx}{4} \left(1-\frac{t}{\cQ^2}\right) +
\frac{4\Bx(1-\Bx) + \epsilon^2}{2\sqrt{1+\epsilon^2}}\frac{t^\prime}{\cQ^2}
\right)
\Bigg\},
\nonumber\\
\phantom{\frac{a}{b}}
\nonumber\\
C_{++}^{\rm unp} (n \!\!\!&=&\!\!\! 2)
= \frac{8(2-y)\left(1-y-\frac{\epsilon^2}{4} y^2 \right)}{(1+\epsilon^2)^{2}}
\Bigg\{
\frac{2 \epsilon^2}{\sqrt{1 + \epsilon^2}(1+\sqrt{1+\epsilon ^2})} \frac{{\widetilde K}^2}{\cQ^2}
\nonumber\\
&&\qquad\qquad\qquad\qquad\qquad\qquad\qquad\qquad
+
\frac{\Bx t\, t^{\prime}}{\cQ^4}
\left(
1-\Bx -\frac{\sqrt{1+\epsilon ^2}-1}{2} + \frac{\epsilon^2}{2\Bx}
\right)
\Bigg\}
\, ,
\nonumber \\
 C^{ {\rm unp},V}_{++} (n \!\!\!&=&\!\!\! 2)=
\frac{8(2-y) \left(1-y-\frac{\epsilon^2}{4} y^2 \right) }{(1+\epsilon ^2)^{2}} \frac{ \Bx t}{\cQ^2}
\Bigg\{
\frac{4{\widetilde K}^2}{\sqrt{1+\epsilon ^2}\cQ^2} +
\frac{1+\sqrt{1+\epsilon ^2}-2\Bx}{2} \left(1+ \frac{t}{\cQ^2}\right) \frac{t^\prime}{\cQ^2}
\Bigg\},
\nonumber\\
C^{{\rm unp},A}_{++} (n \!\!\!&=&\!\!\! 2)=
\frac{4(2-y)\left(1-y-\frac{\epsilon^2}{4} y^2 \right)}{(1+\epsilon^2)^{2}} \frac{t}{\cQ^2}
\Bigg\{
\frac{4(1-2\Bx){\widetilde K}^2}{\sqrt{1+\epsilon^2}\cQ^2}
-\left( 3-\sqrt{1+\epsilon^2}-2\Bx+\frac{\epsilon^2}{\Bx} \right) \frac{\Bx t^\prime}{\cQ^2}
\Bigg\},
\nonumber\\
\phantom{\frac{a}{b}}
\nonumber\\
C_{++}^{\rm unp} (n \!\!\!&=&\!\!\! 3)
= -8 K \left(1-y-\frac{\epsilon^2}{4} y^2 \right)
\frac{\sqrt{1+\epsilon ^2}-1}{(1+\epsilon^2)^{5/2}}
\left\{
(1 - \Bx) \frac{t}{{\cal Q}^2}
+ \frac{\sqrt{1+\epsilon ^2}-1}{2} \left( 1 + \frac{t}{{\cal Q}^2} \right)
\right\}
\, ,
\nonumber\\
C^{ {\rm unp},V}_{++} (n \!\!\!&=&\!\!\! 3)=
-\frac{8 K \left(1-y-\frac{\epsilon^2}{4} y^2 \right)  }{(1+\epsilon ^2)^{5/2}} \frac{ \Bx t}{\cQ^2}
\left\{
\sqrt{1+\epsilon ^2}-1+ \left (1+\sqrt{1+\epsilon ^2}-2\Bx\right)  \frac{t}{\cQ^2}
\right\},
\nonumber\\
C^{{\rm unp},A}_{++} (n \!\!\!&=&\!\!\! 3)=
\frac{16 K \left(1-y-\frac{\epsilon^2}{4} y^2 \right)}{(1+\epsilon^2)^{5/2}} \frac{t\, t^\prime}{\cQ^4}
\left\{
\Bx(1-\Bx)+\frac{\epsilon^2}{4}
\right\},
\nonumber\\
\phantom{\frac{a}{b}}
\nonumber\\
S_{++}^{\rm unp} (n \!\!\!&=&\!\!\! 1)
= \frac{8\lambda K  (2 - y)y}{1 + \epsilon^2}
\Bigg\{ 1+
\frac{1-\Bx+\frac{\sqrt{1+\epsilon^2}-1}{2}}{1+\epsilon^2}\,\frac{t^{\prime}}{\cQ^2}
\Bigg\}
\, ,
\nonumber\\
S^{ {\rm unp},V}_{++} (n \!\!\!&=&\!\!\! 1)=
-\frac{  8 \lambda K (2-y) y}{(1+\epsilon ^2)^{2}} \frac{ \Bx t}{\cQ^2}
\left\{
\sqrt{1+\epsilon ^2}-1+ \left (1+\sqrt{1+\epsilon ^2}-2\Bx\right)  \frac{t}{\cQ^2}
\right\},
\nonumber\\
S^{{\rm unp},A}_{++} (n \!\!\!&=&\!\!\! 1)=
\frac{8 \lambda K (2-y)y}{(1+\epsilon^2)} \frac{t}{\cQ^2}
\Big\{
1-(1-2\Bx)\frac{1+\sqrt{1+\epsilon^2}-2\Bx}{2\sqrt{1+\epsilon^2}}\frac{t^\prime}{\cQ^2}
\Big\},
\nonumber\\
\phantom{\frac{a}{b}}
\nonumber\\
S_{++}^{\rm unp} (n \!\!\!&=&\!\!\! 2)
\nonumber\\
&=&\!\!\!
-\frac{4\lambda \left(1-y-\frac{\epsilon^2}{4} y^2\right)y}{(1+\epsilon^2)^{3/2}}
\left(1+\sqrt{1+\epsilon^2} -2 \Bx\right)
\frac{t^{\prime}}{\cQ^2}
\Bigg\{\frac{\epsilon^2- \Bx (\sqrt{1+\epsilon^2}-1)}{1+\sqrt{\epsilon^2+1} -2 \Bx}
-
\frac{2 \Bx+\epsilon^2}{2\sqrt{1+\epsilon^2}}\,\frac{t^{\prime}}{\cQ^2}
\Bigg\},
\nonumber\\
S^{ {\rm unp},V}_{++} (n \!\!\!&=&\!\!\! 2)=
-\frac{ 4  \lambda \left(1-y-\frac{\epsilon^2}{4} y^2 \right)  y }{(1+\epsilon ^2)^{2}} \frac{ \Bx t}{\cQ^2}
\nonumber\\
&\times&\!\!\!
\left( 1 - (1 - 2 \Bx) \frac{t}{{\cQ}^2} \right)
\left\{
\sqrt{1+\epsilon^2}-1+ \left(1+\sqrt{1+\epsilon ^2}-2\Bx\right)  \frac{t}{\cQ^2}
\right\},
\nonumber\\
S^{{\rm unp},A}_{++} (n \!\!\!&=&\!\!\! 2)=
-\frac{8 \lambda \left(1-y-\frac{\epsilon^2}{4} y^2 \right)y}{(1+\epsilon^2)^{2}} \frac{t\, t^\prime}{\cQ^4}
\nonumber\\
&&\quad\times
\left( 1+\sqrt{1+\epsilon^2}-2\Bx \right)
\Bigg(
1 + \frac{4 (1 - \Bx) \Bx + \epsilon^2}{4 - 2 \Bx + 3 \epsilon^2} \frac{t}{{\cQ}^2}
\Bigg).
\nonumber
\end{eqnarray}

\noindent
\underline{Longitudinal-transverse coefficients:}
\begin{eqnarray}
C_{0 +}^{\rm unp} (n \!\!\!&=&\!\!\! 0)
=
\frac{12 \sqrt{2} K (2-y) \sqrt{1-y-\frac{\epsilon^2}{4} y^2} }{\left(1+\epsilon^2\right)^{5/2}}
\left\{
\epsilon^2+ \frac{2-6 \Bx-\epsilon^2}{3} \frac{t}{\cQ^2}
\right\}
\, ,
\\
C^{\rm unp}_{0 +} (n \!\!\!&=&\!\!\! 1)
=
\frac{8\sqrt{2} \sqrt{1-y-\frac{\epsilon^2}{4} y^2 }}{\left(1+\epsilon ^2\right)^2}
\Bigg\{
(2-y)^2 \frac{t^\prime}{\cQ^2}
\Bigg(
1 - \Bx +  \frac{(1-\Bx) \Bx+ \frac{\epsilon^2}{4}}{\sqrt{1+\epsilon^2}} \frac{t^\prime}{\cQ^2}
\Bigg)
\nonumber\\
&&\qquad
+\frac{1 - y - \frac{\epsilon^2}{4} y^2 }{\sqrt{1+\epsilon^2}} \left(1-(1-2 \Bx) \frac{t}{\cQ^2}\right)
\left(\epsilon ^2- 2 \left(1+\frac{\epsilon ^2}{2 \Bx}\right) \frac{\Bx t }{\cQ^2} \right)
\Bigg\}
\, ,
\nonumber\\
C^{\rm unp}_{0 +} (n \!\!\!&=&\!\!\! 2)
=
-\frac{8 \sqrt{2} K (2-y)\sqrt{1-y-\frac{\epsilon^2}{4} y^2} }{\left(1+\epsilon^2\right)^{5/2}}
\left(1+\frac{\epsilon^2}{2}\right)
\left\{
1 + \frac{1 +\frac{\epsilon^2}{2\Bx}}{1+\frac{\epsilon^2}{2}} \frac{\Bx t}{\cQ^2 }
\right\}
\, , \nonumber\\
S^{\rm unp}_{0 +} (n \!\!\!&=&\!\!\! 1)
=
\frac{8 \lambda \sqrt{2} (2-y) y \sqrt{1-y-\frac{\epsilon^2}{4} y^2}}{
\left(1+\epsilon^2\right)^2} \frac{{\widetilde{K}^2}}{\cQ^2}
\, , \nonumber\\
S^{\rm unp}_{0 +} (n \!\!\!&=&\!\!\! 2)
=
\frac{8\lambda \sqrt{2} K y \sqrt{1-y-\frac{\epsilon^2}{4} y^2}}{\left(1+\epsilon^2\right)^2}
\left(1+\frac{\epsilon^2}{2}\right)
\left\{
1 + \frac{1 +\frac{\epsilon^2}{2\Bx}}{1+\frac{\epsilon^2}{2}} \frac{\Bx t}{\cQ^2 }
\right\}
\, , \nonumber\\
C^{{\rm unp},V}_{0 +} (n \!\!\!&=&\!\!\! 0)
=
\frac{24 \sqrt{2} K (2 - y) \sqrt{1-y-\frac{\epsilon^2}{4} y^2}}{\left(1+\epsilon^2\right)^{5/2}}
\frac{\Bx t}{\cQ^2} \left\{
1 - (1 - 2 \Bx) \frac{t}{\cQ^2}
\right\}
\, , \nonumber\\
C^{{\rm unp},V}_{0 +} (n \!\!\!&=&\!\!\! 1)
=
\frac{16 \sqrt{2} \sqrt{1-y-\frac{\epsilon^2}{4} y^2}}{\left(1+\epsilon^2\right)^{5/2}}
\frac{\Bx t}{\cQ^2}
\left\{
\frac{\widetilde{K}^2 (2 - y)^2}{\cQ^2}
+
\left( 1 - (1 - 2 \Bx) \frac{t}{\cQ^2} \right)^2
\left(
1-y-\frac{\epsilon^2}{4} y^2
\right)
\right\}
\, , \nonumber\\
C^{{\rm unp},V}_{0 +} (n \!\!\!&=&\!\!\! 2)
=
\frac{8 \sqrt{2} K (2 - y) \sqrt{1-y-\frac{\epsilon^2}{4} y^2}}{\left(1+\epsilon^2\right)^{5/2}}
\frac{\Bx t}{\cQ^2}
\left( 1 - (1 - 2 \Bx) \frac{t}{\cQ^2} \right)
\, , \nonumber\\
S^{{\rm unp},V}_{0 +} (n \!\!\!&=&\!\!\! 1)
=
\frac{4 \sqrt{2} \lambda y (2 - y) \sqrt{1-y-\frac{\epsilon^2}{4} y^2}}{\left(1+\epsilon^2\right)^{2}}
\frac{\Bx t}{\cQ^2}
\left\{ 4 (1 - 2 \Bx) \frac{t}{\cQ^2} \left( 1 + \frac{\Bx t}{\cQ^2} \right) + \epsilon^2  \left( 1 + \frac{t}{\cQ^2} \right)^2  \right\}
\, , \nonumber\\
S^{{\rm unp},V}_{0 +} (n \!\!\!&=&\!\!\! 2)
=
- \frac{8 \sqrt{2} \lambda K y \sqrt{1-y-\frac{\epsilon^2}{4} y^2}}{\left(1+\epsilon^2\right)^{2}}
\frac{\Bx t}{\cQ^2}
\left\{ 1 - (1 - 2 \Bx) \frac{t}{\cQ^2} \right\}
\, , \nonumber\\
C^{{\rm unp},A}_{0 +} (n \!\!\!&=&\!\!\! 0)
=
\frac{4 \sqrt{2} K (2 - y) \sqrt{1-y-\frac{\epsilon^2}{4} y^2}}{\left(1+\epsilon^2\right)^{5/2}}
\frac{t}{\cQ^2} (8 - 6 \Bx + 5 \epsilon^2)
\left\{ 1 - \frac{t}{\cQ^2} \frac{2 - 12 \Bx (1 - \Bx) - \epsilon^2}{8 - 6 \Bx + 5 \epsilon^2}
\right\}
\, , \nonumber\\
C^{{\rm unp},A}_{0 +} (n \!\!\!&=&\!\!\! 1)
=
\frac{8 \sqrt{2} \sqrt{1-y-\frac{\epsilon^2}{4} y^2}}{\left(1+\epsilon^2\right)^{5/2}}
\frac{t}{\cQ^2}
\Bigg\{
\frac{\widetilde{K}^2}{\cQ^2}
(1 - 2 \Bx)(2 - y)^2
\nonumber\\
&+&\!\!\!
\left( 1 - (1 - 2 \Bx) \frac{t}{\cQ^2} \right) \left( 1 - y - \frac{y^2 \epsilon^2}{4} \right)
\left(
4 - 2 \Bx + 3 \epsilon^2 + \frac{t}{\cQ^2} (4 \Bx (1 - \Bx) + \epsilon^2 )
\right)
\Bigg\}
\, , \nonumber\\
C^{{\rm unp},A}_{0 +} (n \!\!\!&=&\!\!\! 2)
=
\frac{8 \sqrt{2} K (2 - y) \sqrt{1-y-\frac{\epsilon^2}{4} y^2}}{\left(1+\epsilon^2\right)^{2}}
\frac{t}{\cQ^2}
\Bigg\{
1 - \Bx + \frac{t'}{2 \cQ^2} \frac{4 \Bx (1 - \Bx) + \epsilon^2}{\sqrt{1 + \epsilon^2}}
\Bigg\}
\, , \nonumber\\
S^{{\rm unp},A}_{0 +} (n \!\!\!&=&\!\!\! 1)
=
- \frac{8 \sqrt{2} \lambda y (2 - y) (1 - 2 \Bx) \sqrt{1-y-\frac{\epsilon^2}{4} y^2}}{\left(1+\epsilon^2\right)^{2}}
\frac{t K^2}{\cQ^4}
\, , \nonumber\\
S^{{\rm unp},A}_{0 +} (n \!\!\!&=&\!\!\! 2)
=
- \frac{2 \sqrt{2} \lambda K y \sqrt{1-y-\frac{\epsilon^2}{4} y^2}}{\left(1+\epsilon^2\right)^{2}}
\frac{t}{\cQ^2}
\left(
4 - 4 \Bx + 2 \epsilon^2 + 4 \frac{t}{\cQ^2} (4 \Bx (1 - \Bx) + \epsilon^2)
\right)
\, , \nonumber\\
\end{eqnarray}

\noindent
\underline{Transverse-transverse helicity-flip coefficients:}
\begin{eqnarray}
C^{\rm unp}_{-+} (n \!\!\!&=&\!\!\! 0)
= \frac{8 (2-y)}{\left(1+\epsilon^2\right)^{3/2}}
\Bigg\{
(2-y)^2 \frac{\sqrt{1+\epsilon ^2}-1}{2(1+\epsilon^2)} \frac{\widetilde{K}^2}{\cQ^2}
\\
&&\quad
+\frac{1-y-\frac{\epsilon^2}{4} y^2}{\sqrt{1+\epsilon^2}}
\Bigg(
1 -\Bx-\frac{\sqrt{1+\epsilon^2}-1}{2}+ \frac{\epsilon^2}{2\Bx}
\Bigg) \frac{\Bx t\, t^\prime}{\cQ^4}
\Bigg\}
\, ,
\nonumber\\
C^{\rm unp}_{-+} (n \!\!\!&=&\!\!\! 1)
=
\frac{8 K}{\left(1+\epsilon^2\right)^{3/2}}
\Bigg\{ (2-y)^2 \frac{2-\sqrt{1+\epsilon ^2}}{1+\epsilon ^2}
\Bigg(
\frac{\sqrt{1+\epsilon ^2}-1+\epsilon^2}{2 \left(2-\sqrt{1+\epsilon ^2}\right)}
\left(1-\frac{t}{\cQ^2}\right)-
\frac{\Bx t}{\cQ^2}
\Bigg)
\nonumber\\
&+&\!\!\!
2 \frac{1-y-\frac{\epsilon^2}{4} y^2}{\sqrt{1+\epsilon^2}}
\Bigg(
\frac{1-\sqrt{1+\epsilon^2}+\frac{\epsilon^2}{2}}{2 \sqrt{1+\epsilon^2}}
+
\frac{t}{\cQ^2}
\left( 1-\frac{3 \Bx}{2}+\frac{\Bx+\frac{\epsilon ^2}{2}}{2 \sqrt{1+\epsilon^2}}\right)
\Bigg)
\Bigg\}
\, ,
\nonumber\\
C^{\rm unp}_{-+} (n \!\!\!&=&\!\!\! 2)
=
4(2-y) \left(1 - y - \frac{\epsilon^2}{4} y^2\right)
\frac{1 + \sqrt{1+\epsilon ^2}}{\left(1+\epsilon^2\right)^{5/2}}
\Bigg\{
(2 - 3\Bx) \frac{t}{\cQ^2}
\nonumber\\
&+&\!\!\!
\left(1-2 \Bx+  \frac{2 (1-\Bx)}{1+\sqrt{1+\epsilon^2}}\right)\frac{\Bx t^2}{\cQ^4}
+
\Bigg(
1 + \frac{\sqrt{1+\epsilon^2} + \Bx + (1-\Bx)\frac{t}{\cQ^2}}{1+\sqrt{1+\epsilon^2}} \frac{t}{\cQ^2} \Bigg)
\epsilon^2
\Bigg\}
\, , \nonumber\\
C^{\rm unp}_{-+} (n \!\!\!&=&\!\!\! 3)
=
-8K \left(1-y-\frac{\epsilon^2}{4} y^2 \right)
\frac{1+ \sqrt{1+\epsilon ^2}+\frac{\epsilon^2}{2}}{\left(1+\epsilon^2\right)^{5/2}}
\Bigg\{
1+ \frac{1+\sqrt{1+\epsilon ^2}+\frac{\epsilon^2}{2 \Bx }}{1+\sqrt{1+\epsilon ^2}
+
\frac{\epsilon^2}{2}} \frac{\Bx t}{\cQ^2} \Bigg\}
\, ,
\nonumber\\
S^{\rm unp}_{-+} (n \!\!\!&=&\!\!\! 1)
=
\frac{4 \lambda K (2-y) y}{\left(1+\epsilon^2\right)^2}
\Bigg\{
1 - \sqrt{1+\epsilon^2}+ 2 \epsilon^2 - 2 \left( 1 + \frac{\sqrt{1+\epsilon^2}-1}{2\Bx} \right) \frac{\Bx t}{\cQ^2}
\Bigg\}\,,
\nonumber\\
S^{\rm unp}_{-+} (n \!\!\!&=&\!\!\! 2)
= 2\lambda y \left(1-y-\frac{\epsilon^2}{4} y^2 \right)
\frac{1+\sqrt{1+\epsilon^2}}{\left(1+\epsilon^2\right)^2}
\Bigg(\epsilon^2 - 2 \Bigg( 1 +\frac{\epsilon^2}{2\Bx} \Bigg) \frac{\Bx t}{\cQ^2} \Bigg)
\nonumber\\
&&\qquad
\times\Bigg\{1+
\frac{\sqrt{1+\epsilon^2}-1+ 2\Bx}{1+\sqrt{1+\epsilon^2}}
\frac{t}{\cQ^2} \Bigg\}
\, , \nonumber\\
C^{{\rm unp},V}_{-+} (n \!\!\!&=&\!\!\! 0)
= \frac{4 (2 - y)}{(1 + \epsilon^2)^{5/2}} \frac{\Bx t}{\cQ^2}
\Bigg\{
\frac{2 \widetilde{K}^2}{\cQ^2}
\left(
2 - 2 y + y^2 + \frac{y^2 \epsilon^2}{2}
\right)
\nonumber\\
&-&\!\!\!
\left(
1 - (1 - 2 \Bx) \frac{t}{\cQ^2}
\right)
\left(
1 - y -  \frac{y^2 \epsilon^2}{4}
\right)
\left(
\sqrt{1 + \epsilon^2} - 1 + \left( \sqrt{1 + \epsilon^2} + 1 - 2 \Bx \right) \frac{t}{\cQ^2}
\right)
\Bigg\}
\, . \nonumber\\
C^{{\rm unp},V}_{-+} (n \!\!\!&=&\!\!\! 1)
= \frac{8 K}{(1 + \epsilon^2)^{5/2}} \frac{\Bx t}{\cQ^2}
\Bigg\{
2
\left(
1 - (1 - 2 \Bx) \frac{t}{\cQ^2}
\right)
\left(
2 - 2 y + y^2 + \frac{y^2 \epsilon^2}{2}
\right)
\nonumber\\
&+&\!\!\!
\left(
1 - y -  \frac{y^2 \epsilon^2}{4}
\right)
\left(
3 - \sqrt{1 + \epsilon^2} - \left( 3 (1 - 2 \Bx) + \sqrt{1 + \epsilon^2} \right) \frac{t}{\cQ^2}
\right)
\Bigg\}
\, , \nonumber\\
C^{{\rm unp},V}_{-+} (n \!\!\!&=&\!\!\! 2)
= \frac{4 (2 - y) \left( 1 - y - \frac{y^2 \epsilon^2}{4} \right)}{(1 + \epsilon^2)^{5/2}} \frac{\Bx t}{\cQ^2}
\Bigg\{
4 \frac{\widetilde{K}^2}{\cQ^2}
\nonumber\\
&+&\!\!\!
1 + \sqrt{1 + \epsilon^2}
+
\frac{t}{\cQ^2}
\left(
(1 - 2 \Bx)
\left(
1 - 2 \Bx - \sqrt{1 + \epsilon^2}
\right) \frac{t}{\cQ^2}
-
2 + 4 \Bx + 2 \Bx \sqrt{1 + \epsilon^2}
\right)
\Bigg\}
\, , \nonumber\\
C^{{\rm unp},V}_{-+} (n \!\!\!&=&\!\!\! 3)
= \frac{8 K \left( 1 - y - \frac{y^2 \epsilon^2}{4} \right)}{(1 + \epsilon^2)^{5/2}} \frac{\Bx t}{\cQ^2}
\left(
1 + \sqrt{1 + \epsilon^2}
\right)
\Bigg\{
1 -
\frac{t}{\cQ^2}
\frac{1 - 2 \Bx - \sqrt{1 + \epsilon^2}}{1 + \sqrt{1 + \epsilon^2}}
\Bigg\}
\, , \nonumber\\
S^{{\rm unp},V}_{-+} (n \!\!\!&=&\!\!\! 1)
= \frac{8 \lambda K y (2 - y)}{(1 + \epsilon^2)^{2}} \frac{\Bx t}{\cQ^2}
\left(
1 + \sqrt{1 + \epsilon^2}
\right)
\Bigg\{
1 -
\frac{t}{\cQ^2}
\frac{1 - 2 \Bx - \sqrt{1 + \epsilon^2}}{1 + \sqrt{1 + \epsilon^2}}
\Bigg\}
\, , \nonumber\\
S^{{\rm unp},V}_{-+} (n \!\!\!&=&\!\!\! 2)
= \frac{4 \lambda y \left( 1 - y - \frac{y^2 \epsilon^2}{4} \right)}{(1 + \epsilon^2)^{2}} \frac{\Bx t}{\cQ^2}
\nonumber\\
&\times&\!\!\!
\left(
1 + \sqrt{1 + \epsilon^2}
\right)
\left(
1 - (1 - 2 \Bx) \frac{t}{\cQ^2}
\right)
\Bigg\{
1 -
\frac{t}{\cQ^2}
\frac{1 - 2 \Bx - \sqrt{1 + \epsilon^2}}{1 + \sqrt{1 + \epsilon^2}}
\Bigg\}
\, , \nonumber\\
C^{{\rm unp},A}_{-+} (n \!\!\!&=&\!\!\! 0)
= \frac{4 (2 - y)}{(1 + \epsilon^2)^{2}} \frac{t}{\cQ^2}
\Bigg\{
\frac{t'}{\cQ^2}
\left(
1 - y - \frac{y^2 \epsilon^2}{4}
\right)
\left(
2 \Bx^2 - \epsilon^2 - 3 \Bx + \Bx \sqrt{1 + \epsilon^2}
\right)
\nonumber\\
&+&\!\!\!
\frac{\widetilde{K}^2}{\cQ^2 \sqrt{1 + \epsilon^2}}
\left(
4 - 2 \Bx (2 - y)^2 - 4 y + y^2 - (1 + \epsilon^2)^{3/2}
\right)
\Bigg\}
\, , \nonumber\\
C^{{\rm unp},A}_{-+} (n \!\!\!&=&\!\!\! 1)
= \frac{4 K}{(1 + \epsilon^2)^{5/2}} \frac{t}{\cQ^2}
\Bigg\{
\left( 2 - 2 y + y^2 + \frac{y^2 \epsilon^2}{2} \right)
\nonumber\\
&\times&\!\!\!
\left(
5 - 4 \Bx + 3 \epsilon^2 - \sqrt{1  +\epsilon^2}
-
\frac{t}{\cQ^2}
\left(
1 - \epsilon^2 - \sqrt{1 + \epsilon^2}
-
2 \Bx (4 - 4 \Bx - \sqrt{1 + \epsilon^2})
\right)
\right)
\nonumber\\
&+&\!\!\!
\left(1 - y - \frac{y^2 \epsilon^2}{4} \right)
\nonumber\\
&\times&\!\!\!
\left(
8 + 5 \epsilon^2 - 6 \Bx + 2 \Bx \sqrt{1 + \epsilon^2}
-
\frac{t}{\cQ^2}
\left(
2 - \epsilon^2 + 2 \sqrt{1 + \epsilon^2}
-
4 \Bx (3 - 3 \Bx + \sqrt{1 + \epsilon^2})
\right)
\right)
\Bigg\}
\, , \nonumber\\
C^{{\rm unp},A}_{-+} (n \!\!\!&=&\!\!\! 2)
= \frac{16 (2 - y) \left( 1 - y - \frac{y^2 \epsilon^2}{4} \right)}{(1 + \epsilon^2)^{3/2}} \frac{t}{\cQ^2}
\Bigg\{
\frac{\widetilde{K}^2}{\cQ^2}
\frac{1 - 2 \Bx}{1 + \epsilon^2}
\nonumber\\
&-&\!\!\!
\frac{1 - \Bx}{4 \Bx (1 - \Bx) + \epsilon^2}
\left(
2 \Bx^2 - \epsilon^2 - 3 \Bx - \Bx \sqrt{1 + \epsilon^2}
\right)
-
\frac{t'}{\cQ^2}
\frac{2 \Bx^2 - \epsilon^2 - 3 \Bx - \Bx \sqrt{1 + \epsilon^2}}{4 \sqrt{1 + \epsilon^2}}
\Bigg\}
\, , \nonumber\\
C^{{\rm unp},A}_{-+} (n \!\!\!&=&\!\!\! 3)
= \frac{16K \left( 1 - y - \frac{y^2 \epsilon^2}{4} \right)}{(1 + \epsilon^2)^{2}} \frac{t}{\cQ^2}
\Bigg\{
1 - \Bx + \frac{t'}{\cQ^2} \frac{4 \Bx (1 - \Bx) + \epsilon^2}{4 \sqrt{1 + \epsilon^2}}
\Bigg\}
\, , \nonumber\\
S^{{\rm unp},A}_{-+} (n \!\!\!&=&\!\!\! 1)
= \frac{4 \lambda K y (2 - y)}{(1 + \epsilon^2)^{2}} \frac{t}{\cQ^2}
\Bigg\{
3 + 2 \epsilon^2
\nonumber\\
&+&\!\!\!
\sqrt{1 + \epsilon^2} - 2 \Bx - 2 \Bx \sqrt{1 + \epsilon^2}
-
\frac{t}{\cQ^2} (1 - 2 \Bx)
\left( 1 - 2 \Bx - \sqrt{1 + \epsilon^2} \right)
\Bigg\}
\, , \nonumber\\
S^{{\rm unp},A}_{-+} (n \!\!\!&=&\!\!\! 2)
= \frac{2 \lambda\left( 1 - y - \frac{y^2 \epsilon^2}{4} \right)}{(1 + \epsilon^2)^{2}} \frac{t}{\cQ^2}
\left(
4 - 2 \Bx + 3 \epsilon^2 + \frac{t}{\cQ^2}
\left( 4 \Bx (1 - \Bx) + \epsilon^2 \right)
\right)
\nonumber\\
&\times&\!\!\!
\left(
1 + \sqrt{1 + \epsilon^2}
-
\frac{t}{\cQ^2}
\left( 1 - 2 \Bx - \sqrt{1 + \epsilon^2} \right)
\right)
\, .\nonumber
\end{eqnarray}

\subsection{Longitudinally polarized target}

In the helicity dependent contribution of a longitudinal polarized target the third even harmonic vanishes, i.e.,
$$ C^{\rm LP}_{ab} (n=3)=C^{{\rm LP},V}_{ab} (n=3)=C^{{\rm LP},A}_{ab} (n=3)=0, $$  and will be not listed.

\noindent
\underline{Conserved photon-helicity coefficients:}

\begin{eqnarray}
C_{++}^{\rm LP} (n \!\!\!&=&\!\!\! 0)
=  -\frac{4 \lambda  \Lambda y \left(1+\sqrt{1+\epsilon ^2}\right)  }{(1+\epsilon ^2)^{5/2}}
\Bigg\{
(2-y)^2 \frac{{\widetilde K}^2}{\cQ^2}
+
\left(1-y+\frac{\epsilon^2}{4}y^2\right)
\\
&&\hspace{4.5cm}
\times
\left( \frac{\Bx t}{{\cQ}^2}-\left(1-\frac{t}{{\cQ}^2}\right) \frac{\epsilon ^2}{2} \right)
\left(1+  \frac{\sqrt{1+\epsilon ^2}-1+2\Bx}{1+\sqrt{1+\epsilon ^2}} \frac{t}{\cQ^2}\right)
\Bigg\},
\nonumber\\
C_{++}^{\rm LP,V} (n \!\!\!&=&\!\!\! 0)
=
\frac{4  \lambda  \Lambda y \left(1+\sqrt{1+\epsilon ^2}\right) }{\left(1+\epsilon ^2\right)^{5/2}} \frac{t}{\cQ^2}
\Bigg\{
(2-y)^2\,
\frac{1+\sqrt{1+\epsilon ^2} -2 \Bx}{1+\sqrt{1+\epsilon ^2}}\,\frac{{\widetilde K}^2}{\cQ^2}
+ \left(1 -y -\frac{\epsilon ^2}{4} y^2\right)
\nonumber\\
&&\times \left(2-\Bx+\frac{3 \epsilon ^2}{2} \right)
\left(1+\frac{4 (1-\Bx) \Bx+\epsilon ^2}{4-2 \Bx+ 3 \epsilon ^2} \frac{t}{\cQ^2} \right)
\left(1+\frac{\sqrt{1+\epsilon ^2}-1 + 2 \Bx}{1+\sqrt{1+\epsilon ^2}}  \frac{t}{\cQ^2} \right)
\Bigg\},
\nonumber\\
C_{++}^{\rm LP,A} (n \!\!\!&=&\!\!\! 0)
= \frac{4  \lambda  \Lambda y }{\left(1+\epsilon ^2\right)^{5/2}}
\frac{\Bx t}{\cQ^2}
\Bigg\{
2 (2-y)^2 \frac{{\widetilde K}^2}{\cQ^2} + \left(1 -y -\frac{\epsilon ^2}{4} y^2\right)
(1+\sqrt{1+\epsilon ^2})
\left(1-(1-2 \Bx)\frac{t}{\cQ^2} \right)
\nonumber\\
&&\hspace{8.5cm}\times
\left(1+\frac{\sqrt{1+\epsilon ^2}-1+2 \Bx}{1+\sqrt{1+\epsilon ^2}}\frac{t}{\cQ^2}\right)
\Bigg\},
\nonumber\\
\phantom{\frac{a}{b}}
\nonumber\\
C_{++}^{\rm LP} (n \!\!\!&=&\!\!\! 1)
=
-\frac{4 \lambda  \Lambda  K y (2-y)}{\left(1+\epsilon ^2\right)^{5/2}}
(1+\sqrt{1+\epsilon ^2}-\epsilon^2)
\Bigg\{
1- \left(1-2\Bx
\frac{2+\sqrt{1+\epsilon ^2}}{1+\sqrt{1+\epsilon ^2}-\epsilon ^2}\right)
\frac{t}{\cQ^2}
\Bigg\},
\nonumber\\
C_{++}^{\rm LP,V} (n \!\!\!&=&\!\!\! 1)
= \frac{8  \lambda  \Lambda K (2-y) y }{\left(1+\epsilon ^2\right)^{2}}
\left( \sqrt{1+\epsilon ^2}+2(1- \Bx) \right) \frac{t}{\cQ^2}
\Bigg\{
1-\frac{1+\frac{1-\epsilon ^2}{\sqrt{1+\epsilon ^2}}-2\Bx \left(1+\frac{4 (1-\Bx)}{\sqrt{1+\epsilon ^2}}\right)}{2 \left(\sqrt{1+\epsilon ^2}+2 (1-\Bx)\right)}\frac{t^\prime}{\cQ^2}
\Bigg\},
\nonumber\\
C_{++}^{\rm LP,A} (n \!\!\!&=&\!\!\! 1)
=  \frac{16  \lambda  \Lambda K (2-y) y }{\left(1+\epsilon ^2\right)^{5/2}} \frac{\Bx t}{\cQ^2}\left(1-(1-2\Bx)\frac{t}{\cQ^2}\right),
\nonumber\\
\phantom{\frac{a}{b}}
\nonumber\\
C_{++}^{\rm LP} (n \!\!\!&=&\!\!\! 2)
=-\frac{4  \lambda  \Lambda y \left(1 -y -\frac{\epsilon ^2}{4} y^2\right) }{\left(1+\epsilon ^2\right)^{5/2}}
\left( \frac{\Bx t}{{\cQ}^2}-\left(1-\frac{t}{{\cQ}^2}\right) \frac{\epsilon ^2}{2} \right)
\nonumber\\
&&\hspace{7cm}\times
\left\{
1-\sqrt{1+\epsilon ^2}
- \left(1+\sqrt{1+\epsilon ^2}-2 \Bx\right)\frac{t}{\cQ^2}
\right\},
\nonumber\\
C_{++}^{\rm LP,V} (n \!\!\!&=&\!\!\! 2)
= -\frac{2 \lambda  \Lambda y \left(1 -y -\frac{\epsilon ^2}{4} y^2\right)  }{\left(1+\epsilon ^2\right)^{5/2}}
\left( 4-2 \Bx+3 \epsilon^2 \right) \frac{t}{{\cQ}^2} \left(1+\frac{4 (1-\Bx) \Bx+\epsilon ^2}{4-2 \Bx+3 \epsilon ^2} \frac{t}{{\cQ}^2} \right)
\nonumber\\
&&\hspace{7cm}\times\left\{\sqrt{1+\epsilon ^2}-1 +\left(1+\sqrt{1+\epsilon ^2}-2 \Bx\right)    \frac{t}{{\cQ}^2}\right\},
\nonumber\\
C_{++}^{\rm LP,A} (n \!\!\!&=&\!\!\! 2)
= \frac{4  \lambda  \Lambda y \left(1 -y -\frac{\epsilon ^2}{4} y^2\right) }{\left(1+\epsilon ^2\right)^{5/2}}
\frac{\Bx t}{{\cQ}^2}\left(1-(1-2\Bx) \frac{t}{\cQ^2}\right)
\nonumber\\
&&\hspace{7cm}\times
\Bigg\{
1-\sqrt{1+\epsilon ^2}-\left(1+\sqrt{1+\epsilon ^2}-2\Bx\right)\frac{t}{\cQ^2}
\Bigg\},
\nonumber\\
\phantom{\frac{a}{b}}
\nonumber\\
S_{++}^{\rm LP} (n \!\!\!&=&\!\!\! 1)
=
\frac{4 \Lambda K \left(2-2 y+y^2+ \frac{\epsilon ^2}{2}  y^2 \right)  }{\left(1+\epsilon ^2\right)^3}
(1+\sqrt{1+\epsilon ^2})
\left\{
2 \sqrt{1+\epsilon ^2}-1+
\frac{1+\sqrt{1+\epsilon ^2}-2 \Bx}{1+\sqrt{1+\epsilon ^2}} \frac{t}{\cQ^2}
\right\}
\nonumber\\
&&\quad+
\frac{8 K  \Lambda  \left(1-y-\frac{\epsilon ^2}{4} y^2 \right)}{\left(1+\epsilon ^2\right)^3}
\Bigg\{
\frac{3 \epsilon ^2}{2} +
\left(1-\sqrt{1+\epsilon ^2}-\frac{\epsilon ^2}{2}- \Bx \left(3-\sqrt{1+\epsilon ^2}\right)\right)\frac{t}{\cQ^2}
\Bigg\},
\nonumber\\
S_{++}^{\rm LP,V} (n \!\!\!&=&\!\!\! 1)
= \frac{8 \Lambda K \left(2-2 y+y^2+ \frac{\epsilon ^2}{2}  y^2 \right)  }{\left(1+\epsilon ^2\right)^2} \frac{t}{\cQ^2}
\Bigg\{
1-\frac{(1-2 \Bx) \left(1+\sqrt{1+\epsilon^2}-2 \Bx\right)}{2(1+\epsilon ^2)}\frac{t^\prime}{\cQ^2}
\Bigg\}
\nonumber\\
&&\quad+ \frac{32 \Lambda K  \left(1-y-\frac{\epsilon ^2}{4} y^2 \right) }{\left(1+\epsilon ^2\right)^3}
\left(1- \frac{3+\sqrt{1+\epsilon^2}}{4}  \Bx + \frac{5 \epsilon^2}{8}\right) \frac{t}{\cQ^2}
\nonumber\\
&&\hspace{3cm}\times
\Bigg\{
1-\frac{1- \sqrt{1+\epsilon^2}+\frac{\epsilon^2}{2}-2 \Bx \left(3 (1-\Bx)-\sqrt{1+\epsilon^2}\right)}{4- \Bx \left(\sqrt{1+\epsilon^2}+3\right)+\frac{5 \epsilon^2}{2}}
\frac{t}{\cQ^2}
\Bigg\},
\nonumber\\
S_{++}^{\rm LP,A} (n \!\!\!&=&\!\!\! 1)
=
- \frac{8 \Lambda K \left(2-2 y+y^2+ \frac{\epsilon ^2}{2}  y^2 \right)  }{\left(1+\epsilon ^2\right)^3} \frac{\Bx t}{\cQ^2}
\Bigg\{
\sqrt{1+\epsilon^2}-1 +(1+\sqrt{1+\epsilon^2}-2\Bx)\frac{t}{\cQ^2}
\Bigg\}
\nonumber\\
&&\qquad+ \frac{8 \Lambda K  \left(1-y-\frac{\epsilon ^2}{4} y^2 \right) }{\left(1+\epsilon ^2\right)^3}
(3+\sqrt{1+\epsilon^2}) \frac{\Bx t}{\cQ^2}
\Bigg\{
1- \frac{3-\sqrt{1+\epsilon^2}-6 \Bx}{3+\sqrt{1+\epsilon^2}}
\frac{t}{\cQ^2}
\Bigg\},
\nonumber\\
\phantom{\frac{a}{b}}
\nonumber\\
S_{++}^{\rm LP} (n \!\!\!&=&\!\!\! 2)
=
-\frac{4\Lambda  (2-y) \left(1-y-\frac{\epsilon ^2}{4} y^2 \right) }{\left(1+\epsilon ^2\right)^{5/2}}
\nonumber\\
&\times&\!\!\!
\Bigg\{
\frac{4 {\widetilde K}^2}{\sqrt{1+\epsilon ^2}\cQ^2 }
(1+\sqrt{1+\epsilon ^2}-2 \Bx)
\left(1+\sqrt{1+\epsilon ^2}   +\frac{\Bx t}{\cQ^2}\right)\frac{t^\prime}{\cQ^2}
\Bigg\},
\nonumber\\
S_{++}^{\rm LP,V} (n \!\!\!&=&\!\!\! 2)
= \frac{4 \Lambda  (2-y) \left(1-y-\frac{\epsilon ^2}{4} y^2 \right) }{\left(1+\epsilon ^2\right)^{5/2}} \frac{t}{\cQ^2}
\Bigg\{
\frac{4 (1-2\Bx) {\widetilde K}^2}{\sqrt{1+\epsilon ^2}\cQ^2 }
-
\left(
3-\sqrt{1+\epsilon ^2} -2 \Bx+ \frac{\epsilon^2}{\Bx} \right)
\frac{\Bx t^\prime}{\cQ^2}
\Bigg\},
\nonumber\\
S_{++}^{\rm LP,A} (n \!\!\!&=&\!\!\! 2)
= \frac{4 \Lambda  (2-y) \left(1-y-\frac{\epsilon ^2}{4} y^2 \right) }{\left(1+\epsilon ^2\right)^{5/2}} \frac{\Bx t}{\cQ^2}
\Bigg\{
\frac{4 {\widetilde K}^2}{\cQ^2}
-
\left( 1+\sqrt{1+\epsilon^2}-2\Bx \right)
\left(1- \frac{(1-2\Bx) t}{\cQ^2}\right)\frac{t^\prime}{\cQ^2}
\Bigg\},
\nonumber\\
\phantom{\frac{a}{b}}
\nonumber\\
S_{++}^{\rm LP} (n \!\!\!&=&\!\!\! 3)
=
-\frac{4\Lambda K \left(1-y-\frac{\epsilon ^2}{4} y^2 \right) }{\left(1+\epsilon ^2\right)^{3}}
\frac{1+\sqrt{1+\epsilon ^2}-2\Bx}{1+\sqrt{1+\epsilon^2}}
\frac{\epsilon^2 t^\prime}{\cQ^2} ,
\nonumber\\
S_{++}^{\rm LP,V} (n \!\!\!&=&\!\!\! 3)
= \frac{4 \Lambda K \left(1-y-\frac{\epsilon ^2}{4} y^2 \right) }{\left(1+\epsilon ^2\right)^{3}}
 \left(4 (1-\Bx)\Bx + \epsilon^2 \right)  \frac{t\, t^\prime}{\cQ^4},
\nonumber\\
S_{++}^{\rm LP,A} (n \!\!\!&=&\!\!\! 3)
= - \frac{8 \Lambda K \left(1-y-\frac{\epsilon ^2}{4} y^2 \right) }{\left(1+\epsilon ^2\right)^{3}}
\left( 1+\sqrt{1+\epsilon ^2}-2\Bx \right) \frac{\Bx t\, t^\prime}{\cQ^4}
\, .
\nonumber
\end{eqnarray}

\noindent
\underline{photon helicity-flip amplitudes by one unit:}
\begin{eqnarray}
C_{0+}^{\rm LP} (n \!\!\!&=&\!\!\! 0)
=  \frac{8 \sqrt{2}  \lambda \Lambda K (1 - \Bx) y \sqrt{1 - y - \frac{y^2 \epsilon^2}{4}}}{(1+\epsilon ^2)^2}
\frac{t}{\cQ^2}
\, , \nonumber\\
C_{0+}^{\rm LP} (n \!\!\!&=&\!\!\! 1)
=  - \frac{8 \sqrt{2}  \lambda \Lambda K  y (1 - y) \sqrt{1 - y - \frac{y^2 \epsilon^2}{4}}}{(1+\epsilon ^2)^2}
\frac{\widetilde{K}^2}{\cQ^2}
\, , \nonumber\\
C_{0+}^{\rm LP} (n \!\!\!&=&\!\!\! 2)
=  - \frac{8 \sqrt{2}  \lambda \Lambda K y \sqrt{1 - y - \frac{y^2 \epsilon^2}{4}}}{(1+\epsilon ^2)^2}
\left( 1 + \frac{\Bx t}{\cQ^2} \right)
\, , \nonumber\\
S_{0+}^{\rm LP} (n \!\!\!&=&\!\!\! 1)
=  \frac{8 \sqrt{2} \Lambda \sqrt{1 - y - \frac{y^2 \epsilon^2}{4}}}{(1+\epsilon ^2)^{5/2}}
\Bigg\{
\frac{\widetilde{K}^2}{\cQ^2} (2 - y)^2
\nonumber\\
&+&\!\!\!
\left(
1 + \frac{t}{\cQ^2}
\right)
\left(
1 - y - \frac{y^2 \epsilon^2}{4}
\right)
\left(
2 \frac{\Bx t}{\cQ^2} - \left( 1 - \frac{t}{\cQ^2} \right) \epsilon^2
\right)
\Bigg\}
\, , \nonumber\\
S_{0+}^{\rm LP} (n \!\!\!&=&\!\!\! 2)
=  \frac{8 \sqrt{2} \Lambda K (2 - y) \sqrt{1 - y - \frac{y^2 \epsilon^2}{4}}}{(1+\epsilon ^2)^{5/2}}
\left(
1 + \frac{\Bx t}{\cQ^2}
\right)
\, , \nonumber\\
C_{0+}^{{\rm LP}, V} (n \!\!\!&=&\!\!\! 0)
=  \frac{8 \sqrt{2} \lambda \Lambda K y \sqrt{1 - y - \frac{y^2 \epsilon^2}{4}}}{(1+\epsilon ^2)^2}
\frac{t}{\cQ^2}
\left(
\Bx - \frac{t}{\cQ^2} (1 - 2\Bx)
\right)
\, , \nonumber\\
C_{0+}^{{\rm LP}, V} (n \!\!\!&=&\!\!\! 1)
=  \frac{8 \sqrt{2} \lambda \Lambda y (2 - y) \sqrt{1 - y - \frac{y^2 \epsilon^2}{4}}}{(1+\epsilon ^2)^2}
\frac{t \widetilde{K}^2}{\cQ^4}
\, , \nonumber\\
C_{0+}^{{\rm LP}, V} (n \!\!\!&=&\!\!\! 2)
=  \frac{8 \sqrt{2} \lambda \Lambda K y  (1 - \Bx) \sqrt{1 - y - \frac{y^2 \epsilon^2}{4}}}{(1+\epsilon ^2)^2}
\frac{t}{\cQ^2}
\, , \nonumber\\
S_{0+}^{{\rm LP}, V} (n \!\!\!&=&\!\!\! 1)
=  - \frac{8 \sqrt{2} \Lambda \sqrt{1 - y - \frac{y^2 \epsilon^2}{4}}}{(1+\epsilon ^2)^{5/2}}
\frac{t}{\cQ^2}
\Bigg\{
\frac{\widetilde{K}^2}{\cQ^2} (2 - y)^2
\nonumber\\
&+&\!\!\!
\left( 1 + \frac{t}{\cQ^2} \right) \left( 1 - y - \frac{y^2 \epsilon^2}{4} \right)
\left( 4 - 2 \Bx + 3 \epsilon^2 + \frac{t}{\cQ^2} (4 \Bx (1 - \Bx) + \epsilon^2) \right)
\Bigg\}
\, , \nonumber\\
S_{0+}^{{\rm LP}, V} (n \!\!\!&=&\!\!\! 2)
=  - \frac{8 \sqrt{2} \Lambda K (2 - y) (1 - \Bx) \sqrt{1 - y - \frac{y^2 \epsilon^2}{4}}}{(1+\epsilon ^2)^{5/2}}
\frac{t}{\cQ^2}
\, , \nonumber\\
C_{0+}^{{\rm LP}, A} (n \!\!\!&=&\!\!\! 0)
=  - \frac{8 \sqrt{2} \lambda \Lambda K y \sqrt{1 - y - \frac{y^2 \epsilon^2}{4}}}{(1+\epsilon ^2)^2}
\frac{\Bx t}{\cQ^2} \left( 1 + \frac{t}{\cQ^2} \right)
\, , \nonumber\\
C_{0+}^{{\rm LP}, A} (n \!\!\!&=&\!\!\! 2)
=  \frac{8 \sqrt{2} \lambda \Lambda K y \sqrt{1 - y - \frac{y^2 \epsilon^2}{4}}}{(1+\epsilon ^2)^2}
\frac{\Bx t}{\cQ^2} \left( 1 + \frac{t}{\cQ^2} \right)
\, , \nonumber\\
S_{0+}^{{\rm LP}, A} (n \!\!\!&=&\!\!\! 1)
=  - \frac{16 \sqrt{2} \Lambda \left(1 - y - \frac{y^2 \epsilon^2}{4} \right)^{3/2}}{(1+\epsilon ^2)^{5/2}}
\frac{\Bx t}{\cQ^2} \left( 1 + \frac{t}{\cQ^2} \right)
\left( 1 - (1 - 2 \Bx) \frac{t}{\cQ^2} \right)
\, , \nonumber\\
S_{0+}^{{\rm LP}, A} (n \!\!\!&=&\!\!\! 2)
=  - \frac{8 \sqrt{2} \Lambda K (2 - y) \sqrt{1 - y - \frac{y^2 \epsilon^2}{4}}}{(1+\epsilon ^2)^{5/2}}
\frac{\Bx t}{\cQ^2} \left( 1 + \frac{t}{\cQ^2} \right)
\, . \nonumber
\end{eqnarray}

\noindent
\underline{Photon helicity-flip amplitudes by two units:}
\begin{eqnarray}
C_{-+}^{\rm LP} (n \!\!\!&=&\!\!\! 0)
= \frac{4 \lambda \Lambda y}{(1+\epsilon ^2)^{5/2}}
\Bigg\{
\frac{\widetilde{K}^2}{\cQ^2} (2 - y)^2
\left(
1 - \sqrt{1 + \epsilon^2}
\right)
\nonumber\\
&+&\!\!\!
\frac{1}{2}
\left(
1 - y - \frac{y^2 \epsilon^2}{4}
\right)
\left(
2 \frac{\Bx t}{\cQ^2}
-
\left( 1 - \frac{t}{\cQ^2} \right) \epsilon^2
\right)
\left(
1 - \sqrt{1 + \epsilon^2}
-
\frac{t}{\cQ^2}
\left(
1 - 2 \Bx + \sqrt{1 + \epsilon^2}
\right)
\right)
\Bigg\}
\, , \nonumber\\
C_{-+}^{\rm LP} (n \!\!\!&=&\!\!\! 1)
= \frac{4 \lambda \Lambda K y (2 - y)}{(1+\epsilon ^2)^{5/2}}
\Bigg\{
1 - \epsilon^2 - \sqrt{1 + \epsilon^2}
-
\frac{t}{\cQ^2}
\left(
1 - \epsilon^2 - \sqrt{1 + \epsilon^2}
-
2 \Bx
\left(
2 - \sqrt{1 + \epsilon^2}
\right)
\right)
\Bigg\}
\, , \nonumber\\
C_{-+}^{\rm LP} (n \!\!\!&=&\!\!\! 2)
= - \frac{2 \lambda \Lambda y \left( 1 - y - \frac{y^2 \epsilon^2}{4} \right)}{(1+\epsilon ^2)^{5/2}}
\Bigg\{
\epsilon^2
\left(
1 + \sqrt{1 + \epsilon^2}
\right)
\nonumber\\
&-&\!\!\!
2 \frac{t}{\cQ^2}
\left(
 (1 - \Bx) \epsilon^2 + \Bx \left( 1 + \sqrt{1 + \epsilon^2} \right)
\right)
+
\frac{t^2}{\cQ^4}
(2 \Bx + \epsilon^2)
\left(
1 - 2 \Bx - \sqrt{1 + \epsilon^2}
\right)
\Bigg\}
\, , \nonumber\\
S_{-+}^{\rm LP} (n \!\!\!&=&\!\!\! 1)
= - \frac{4 \Lambda K}{(1+\epsilon ^2)^{3}}
\Bigg\{
(2 - y)^2
\left(
1 + 2 \epsilon^2
-
\sqrt{1 + \epsilon^2}
+
\frac{t}{\cQ^2}
\left(
1 - 2 \Bx - \sqrt{1 + \epsilon^2}
\right)
\right)
\nonumber\\
&-&\!\!\!
\left(
1 - y - \frac{y^2 \epsilon^2}{4}
\right)
\left(
2 + \epsilon^2
-
2 \sqrt{1 + \epsilon^2}
+
\frac{t}{\cQ^2}
\left(
\epsilon^2 - 4 \sqrt{1 + \epsilon^2}
+
2 \Bx (1 + \sqrt{1 + \epsilon^2})
\right)
\right)
\Bigg\}
\, , \nonumber\\
S_{-+}^{\rm LP} (n \!\!\!&=&\!\!\! 2)
= - \frac{4 \Lambda (2 - y) \left( 1 - y - \frac{y^2 \epsilon^2}{4} \right)}{(1+\epsilon ^2)^{3}}
\Bigg\{
\frac{t}{\cQ^2}
\left(
2 + 2 \sqrt{1 + \epsilon^2} + \epsilon^2 \sqrt{1 + \epsilon^2}
-
\Bx \left( 3 - \epsilon^2 + 3 \sqrt{1 + \epsilon^2} \right)
\right)
\nonumber\\
&+&\!\!\!
\frac{t^2}{\cQ^4}
\left(
\epsilon^2 - 2 \Bx^2 (2 + \sqrt{1 + \epsilon^2})
+ \Bx (3 - \epsilon^2 + \sqrt{1 + \epsilon^2})
\right)
+
\epsilon^2 \left( 1 + \sqrt{1 + \epsilon^2} \right)
\Bigg\}
\, , \nonumber\\
S_{-+}^{\rm LP} (n \!\!\!&=&\!\!\! 3)
= \frac{4 \Lambda K \left( 1 - y - \frac{y^2 \epsilon^2}{4} \right)}{(1+\epsilon ^2)^{3}}
\Bigg\{
2 + \epsilon^2 + 2 \sqrt{1 + \epsilon^2}
+
\frac{t}{\cQ^2} \left( \epsilon^2 + 2 \Bx (1 + \sqrt{1 + \epsilon^2}) \right)
\Bigg\}
\, , \nonumber\\
C_{-+}^{\rm LP} (n \!\!\!&=&\!\!\! 0)
= \frac{2 \lambda \Lambda y}{(1+\epsilon ^2)^{5/2}} \frac{t}{\cQ^2}
\Bigg\{
(4 - 2 \Bx + 3 \epsilon^2)
\left(
1 - y - \frac{y^2 \epsilon^2}{4}
\right)
\left(
1 + \frac{t}{\cQ^2}
\frac{4 \Bx (1 - \Bx) + \epsilon^2}{4 - 2 \Bx + 3 \epsilon^2}
\right)
\nonumber\\
&\times&\!\!\!
\left(
\sqrt{1 + \epsilon^2} - 1 + \frac{t}{\cQ^2}
\left(
1 - 2 \Bx + \sqrt{1 + \epsilon^2}
\right)
\right)
+
2 (2 - y)^2
(\sqrt{1 + \epsilon^2} - 1 + 2 \Bx) \frac{\widetilde{K}^2}{\cQ^2}
\Bigg\}
\, , \nonumber\\
C_{-+}^{\rm LP} (n \!\!\!&=&\!\!\! 1)
= - \frac{4 \lambda \Lambda y (2 - y)}{(1+\epsilon ^2)^{5/2}} \frac{t}{\cQ^2}
\Bigg\{
5 - 4 \Bx + 3 \epsilon^2
-
\sqrt{1 + \epsilon^2}
\nonumber\\
&-&\!\!\!
\frac{t}{\cQ^2}
\left(
1 - \epsilon^2 - \sqrt{1+ \epsilon^2}
-
2 \Bx (4 - 4 \Bx - \sqrt{1 + \epsilon^2})
\right)
\Bigg\}
\, , \nonumber\\
C_{-+}^{\rm LP} (n \!\!\!&=&\!\!\! 2)
= - \frac{2 \lambda \Lambda y \left( 1 - y - \frac{y^2 \epsilon^2}{4} \right)}{(1+\epsilon ^2)^{5/2}} \frac{t}{\cQ^2}
\left(
4 - 2 \Bx + 3 \epsilon^2
+ \frac{t}{\cQ^2} (4 \Bx (1 - \Bx) + \epsilon^2)
\right)
\nonumber\\
&\times&\!\!\!
\left(
1 + \sqrt{1 + \epsilon^2} - \frac{t}{\cQ^2} (1 - \sqrt{1 + \epsilon^2} - 2 \Bx)
\right)
\, , \nonumber\\
S_{-+}^{{\rm LP},V} (n \!\!\!&=&\!\!\! 1)
= - \frac{4 \Lambda K}{(1+\epsilon^2)^3}
\frac{t}{\cQ^2}
\Bigg\{
\left( 2 - 2 y + y^2 + \frac{y^2 \epsilon^2}{2} \right)
\nonumber\\
&\times&\!\!\!
\left(
3 + 2 \epsilon^2 + \sqrt{1 + \epsilon^2} - 2 \Bx (1 + \sqrt{1 + \epsilon^2})
-
\frac{t}{\cQ^2}
(1 - 2 \Bx)
(1 - 2 \Bx - \sqrt{1 + \epsilon^2})
\right)
\nonumber\\
&+&\!\!\!
\left(
1 - y - \frac{y^2 \epsilon^2}{4}
\right)
\bigg(
8 + 5 \epsilon^2 - 2 \Bx (3 - \sqrt{1 + \epsilon^2})
\nonumber\\
&&\qquad\qquad\qquad
-
\frac{t}{\cQ^2}
\left(
2 - \epsilon^2 + 2 \sqrt{1 + \epsilon^2}
- 12 \Bx (1 - \Bx) - 4 \Bx \sqrt{1 + \epsilon^2}
\right)
\bigg)
\Bigg\}
\, , \nonumber\\
S_{-+}^{{\rm LP},V} (n \!\!\!&=&\!\!\! 2)
= - \frac{4 \Lambda (2 - y) \sqrt{1 - y - \frac{y^2 \epsilon^2}{4}}}{(1+\epsilon^2)^{5/2}}
\frac{t}{\cQ^2}
\Bigg\{
(2 - \Bx) (1 + \sqrt{1 + \epsilon^2})
\nonumber\\
&+&\!\!\!
\epsilon^2
+
\frac{4 \widetilde{K}^2 (1 - 2 \Bx)}{\cQ^2 \sqrt{1 + \epsilon^2}}
+
\frac{t}{\cQ^2}
\left(
\epsilon^2 + \Bx (3 - 2 \Bx + \sqrt{1 + \epsilon^2})
\right)
\Bigg\}
\, , \nonumber\\
S_{-+}^{{\rm LP},V} (n \!\!\!&=&\!\!\! 3)
= - \frac{4 \Lambda K \left(1 - y - \frac{y^2 \epsilon^2}{4} \right)}{(1+\epsilon^2)^{5/2}}
\frac{t}{\cQ^2}
\Bigg\{
4 - 4 \Bx + \frac{t'}{\cQ^2} \frac{4 \Bx (1 - \Bx) + \epsilon^2}{\sqrt{1 + \epsilon^2}}
\Bigg\}
\, , \nonumber\\
C_{-+}^{{\rm LP},A} (n \!\!\!&=&\!\!\! 0)
=  \frac{4 \lambda \Lambda \Bx y}{(1+\epsilon^2)^{5/2}}
\frac{t}{\cQ^2}
\Bigg\{
2 (2 - y)^2
\left(
(1 - \Bx) \frac{t}{\cQ^2}  \left(1 + \frac{\Bx t}{\cQ^2} \right)
+
\left( 1 + \frac{t}{\cQ^2} \right)^2
\frac{\epsilon^2}{4}
\right)
\nonumber\\
&-&\!\!\!
\left(
1 - y - \frac{y^2 \epsilon^2}{4}
\right)
\left(
1 - (1 - 2 \Bx) \frac{t}{\cQ^2}
\right)
\left(
1 - \sqrt{1 + \epsilon^2}
-
\frac{t}{\cQ^2}
(1 + \sqrt{1 + \epsilon^2} - 2 \Bx)
\right)
\Bigg\}
\, , \nonumber\\
C_{-+}^{{\rm LP},A} (n \!\!\!&=&\!\!\! 1)
=  - \frac{16 \lambda \Lambda \Bx y (2 - y)}{(1+\epsilon^2)^{5/2}}
\frac{t}{\cQ^2}
\left(1 - (1 - 2 \Bx) \frac{t}{\cQ^2} \right)
\, , \nonumber\\
C_{-+}^{{\rm LP},A} (n \!\!\!&=&\!\!\! 2)
=  - \frac{4 \lambda \Lambda \Bx y \left( 1 - y - \frac{y^2 \epsilon^2}{4} \right)}{(1+\epsilon^2)^{5/2}}
\nonumber\\
&\times&\!\!\!
\frac{t}{\cQ^2}
\left(1 - (1 - 2 \Bx) \frac{t}{\cQ^2} \right)
\left\{
1 + \sqrt{1 + \epsilon^2} - \frac{t}{\cQ^2} \left( 1 - \sqrt{1 + \epsilon^2} - 2 \Bx \right)
\right\}
\, , \nonumber\\
S_{-+}^{{\rm LP},A} (n \!\!\!&=&\!\!\! 1)
=  - \frac{8 \Lambda K \left( 2 - 2 y + y^2 + \frac{y^2 \epsilon^2}{2} \right)}{(1+\epsilon^2)^3}
(1 + \sqrt{1 + \epsilon^2})
\frac{\Bx t}{\cQ^2}
\left(
1 -  \frac{t}{\cQ^2} \frac{1 - \sqrt{1 + \epsilon^2} - 2 \Bx}{1 + \sqrt{1 + \epsilon^2}}
\right)
\nonumber\\
&-&\!\!\!
\frac{8 \Lambda K \left( 1 - y + \frac{y^2 \epsilon^2}{4} \right)}{(1+\epsilon^2)^3}
\frac{\Bx t}{\cQ^2}
\left\{
3 - \sqrt{1 + \epsilon^2}
-
\frac{t}{\cQ^2}
\left(
3 + \sqrt{1 + \epsilon^2} - 6 \Bx
\right)
\right\}
\, , \nonumber\\
S_{-+}^{{\rm LP},A} (n \!\!\!&=&\!\!\! 2)
=  - \frac{4 \Lambda (2 - y) \left( 1 - y - \frac{y^2 \epsilon^2}{4} \right)}{(1+\epsilon^2)^3}
\frac{\Bx t}{\cQ^2}
\Bigg\{
1 + 4 \frac{\widetilde{K}^2}{\cQ^2}
\nonumber\\
&+&\!\!\! \sqrt{1 + \epsilon^2}
-
2 \frac{t}{\cQ^2} \left( 1 - 2 \Bx - \Bx \sqrt{1 + \epsilon^2} \right)
-
\frac{t}{\cQ^2} (1 - 2 \Bx) \left( 1 - 2 \Bx  - \sqrt{1 + \epsilon^2} \right)
\Bigg\}
\, , \nonumber\\
S_{-+}^{{\rm LP},A} (n \!\!\!&=&\!\!\! 3)
=  - \frac{8 \Lambda K \left( 1 - y - \frac{y^2 \epsilon^2}{4} \right)}{(1+\epsilon^2)^3}
\frac{\Bx t}{\cQ^2}
\left\{
1 + \sqrt{1 + \epsilon^2}
- \frac{t}{\cQ^2} \left( 1 - 2 \Bx - \sqrt{1 + \epsilon^2} \right)
\right\}
\, .
\end{eqnarray}

\end{document}